\documentclass[nofootinbib,twocolumn,superscriptaddress]{revtex4-2}
\usepackage[utf8]{inputenc}

\usepackage{xcolor,comment}

\usepackage[caption=false]{subfig}
\usepackage{cancel}
\usepackage{tikz}
\usepackage{tikz-3dplot}
\usepackage{amsmath}
\usepackage{amsfonts}
\usepackage{natbib}
\usepackage{graphicx}
\usepackage{lipsum}
\usepackage{layout}
\usepackage{booktabs}

\usepackage{hyperref}
\hypersetup{
    colorlinks,
    linkcolor={blue!80!black},
    citecolor={blue!80!black},
    urlcolor={blue!80!black}
}
\hypersetup{hidelinks}
\usepackage{xifthen}
\usepackage{fontawesome}

\newcommand{\Bell}{\boldsymbol{\ell}}

\newcommand\pdfref[3]{%
    \href{phd://open-paper?id=#1&page=#2}{%
    \textup{[\textbf{\ifthenelse{\isempty{#3}}{here}{#3}}]}}%
}

\makeatletter
\def\l@subsubsection#1#2{}
\makeatother

\begin{document}

\title{\makebox[0pt]{Mapping cluster angular momentum with the rotational kinetic Sunyaev--Zel'dovich effect}}

\author{Jack Y.\ L.\ Kwok}
\email[E-mail:\ ]{ylk26@cam.ac.uk}
\affiliation{DAMTP, Centre of Mathematical Sciences, University of Cambridge, Wilberforce Road, Cambridge CB3 0WA, United Kingdom}
\affiliation{Kavli Institute for Cosmology Cambridge, Madingley Road, Cambridge CB3 0HA, UK}
\author{Bernardita Ried Guachalla}
\affiliation{Kavli Institute for Cosmology Cambridge, Madingley Road, Cambridge CB3 0HA, UK}
\affiliation{Department of Physics, Stanford University, Stanford, CA, USA 94305-4085}
\affiliation{Kavli Institute for Particle Astrophysics and Cosmology, 382 Via Pueblo Mall, Stanford, CA 94305-4060, USA}
\affiliation{SLAC National Accelerator Laboratory, 2575 Sand Hill Road, Menlo Park, California 94025, USA}
\author{{William~R.~Coulton}}
\affiliation{Department of Physics, University of Oxford, Denys Wilkinson Building, Keble Road, Oxford OX1 3RH, UK.}
\author{Selim C.\ Hotinli}
\affiliation{Perimeter Institute for Theoretical Physics, 31 Caroline St N, Waterloo, ON N2L 2Y5, Canada}
\author{Emmanuel Schaan}
\affiliation{Kavli Institute for Particle Astrophysics and Cosmology, 382 Via Pueblo Mall, Stanford, CA 94305-4060, USA}
\affiliation{SLAC National Accelerator Laboratory, 2575 Sand Hill Road, Menlo Park, California 94025, USA}

\date{\today}

\begin{abstract}
	Galaxy clusters carry angular momentum. If the intracluster gas rotates coherently, the kinetic Sunyaev-Zel'dovich (kSZ) effect imprints a dipole on the microwave sky centered on the cluster. 
We show that microwave maps alone can be used to reconstruct a noisy vector field tracing the projected angular momentum of the ionized gas.
We present an optimal multifrequency matched filter (MMF) that isolates the rotational kSZ (rkSZ) signal from the much larger miscentered thermal Sunyaev-Zel'dovich (tSZ) effect, cosmic microwave, and infrared background contributions, improving on both real-space oriented stacking and matched filtering of maps cleaned with full-sky multifrequency weights.
For the rkSZ profile, we introduce a simple model for coherent intracluster gas rotation and validate it against TNG-Cluster and TNG300-1. The model consists of differentially rotating concentric shells with constant equatorial tangential velocity, characterized by a mass- and redshift-indepedent tangential-to-circular velocity ratio of $v_{\rm tan}/V_{c,200c}=0.18$.
Using this model, we forecast single-cluster signal-to-noise (SNR) as a function of mass and redshift for Simons Observatory; cross-correlation with an external catalog of rotation proxies yields a cumulative detection SNR.
For an idealized mass-limited catalog with perfectly known spin directions, the cumulative SNR of our optimal MMF reaches $\approx 53$, $25$, and $6$ when including all clusters above $\log_{10}(M_{200c}/M_\odot)>14$, 14.5, and 15, respectively.
This work provides a practical framework for CMB measurements of intracluster gas rotation, with applications to the rotational contribution to non-thermal pressure in cluster mass calibration, as well as spin alignments and clustering biases.
\end{abstract}

\maketitle

\section{Introduction}

Galaxy groups and clusters are key objects for both astrophysics and cosmology. Their abundance as a function of mass and redshift is sensitive to the amplitude of primordial density fluctuations and to the growth of structure, making them powerful cosmological probes \cite{allen11_cluster,kravtsov12_cluster,planck16_cluster,ghirardini24_cluster,abbott25_cluster,bocquet24_cluster}. Astrophysically, these massive objects are important for studying feedback from active galactic nuclei (AGN) \cite{mcnamara07_feedback,fabian12_feedback,eckert21_feedback,hadzhiyska25_ksz,bigwood26_feedback,siegel26_feedback}, and the influence of environment on galaxy evolution and quenching \cite{gunn72_env,dressler80_env,boselli06_env,boselli22_env,peng10_quenching,wetzel13_quenching}.
While groups and clusters have been studied extensively, comparatively little is observationally known about their rotation, and how angular momentum is structured within them.

According to tidal torque theory (TTT), haloes acquire their angular momenta in the linear regime until they reach maximum expansion, through gravitational torques exerted on
the protohaloes due to tidal fields from neighboring matter inhomogeneities
\cite{hoyle49_ttt,peebles69_ttt,doroshkevich70_ttt,white84_ttt,barnes87_ttt,heavens88_ttt}.
TTT predicts a halo mass scaling of angular momentum amplitude of
$J\propto M^{5/3}$ and a universal (mass- and redshift-independent) log-normal distribution of dark-matter halo spin parameter
\cite{peebles69_ttt,white84_ttt,white94_ttt,catelan96a_ttt,sugerman00_ttt}, which are verified by numerical works \cite{efstathiou79_spinsim,barnes87_ttt, sugerman00_ttt, bullock01_spinsim,porciani02_ttt}.
In Bullock's parametrization of the spin parameter \cite{bullock01_spinsim},
\begin{align}
	\lambda_x \equiv \frac{j_x}{\sqrt{2} R_{200c}V_{c,200c}} \; ,
\end{align}
where $x$ labels the component inside the halo,  $j_x$ is the specific angular momentum (sAM) magnitude of that component within the sphere of radius  $R_{200c}$, and $V_{c,200c}\equiv\sqrt{GM_{200c}/R_{200c}}$ is the circular velocity at  $R_{200c}$. For dark matter haloes in $N$-body simulations, it is found that $\lambda_\text{DM}\sim0.035$ \cite{bullock01_spinsim,maccio07_spinsim,bett07_spinsim}. A constant distribution of $\lambda$ over time also corresponds to the prediction that the average sAM of freshly accreted material is higher than that of their respective haloes \cite{stewart13_angacquisition}.

Radiative simulations find that freshly accreted dark matter and gas carry comparable specific angular momentum (sAM) \cite{kimm11_spinparam}, as expected since both experienced the same tidal torques in the linear regime \cite{fall80_disc,efstathiou83_galformation}. The subsequent evolution of the spin parameters, however, depends on the halo mass. In haloes below the shock-heating mass threshold \cite{birnboim03_shockheat,ocvirk08_bimodelaccretion}, cold filamentary streams penetrate the virial shock and rapidly deliver gas to the inner halo with 2–5 times the sAM of the halo dark matter \cite{kimm11_spinparam,stewart11_coldgas}. Above the threshold, accreted gas is instead shock-heated and virialised, allowing sufficient time for its angular momentum to be redistributed \cite{stewart13_angacquisition}; for galaxy clusters, hydrodynamical simulations find that the intracluster medium (ICM) carries angular momentum, with a median spin parameter $\lambda_\text{gas}\sim 0.04-0.05$, which is comparable but slightly higher than the dark matter spin parameter \cite{ gottloeber07_spinsim,vdbosch02_spinparam,bett10_jdist,baldi17_rot}. 

The angular momentum content of clusters, however, rests on theoretical and numerical works, as observational determination of the gas angular momentum at cluster scales remains limited.
This is relevant for accurate cluster mass estimation, since rotation along with turbulent intra-cluster medium (ICM) motions \cite{lau09_icmmotion} provide non-thermal pressure support; the hydrostatic mass measured from thermal pressure support underestimates the true mass by around 10\% at $R_{200c}$ \cite{eckert17_xcop}. Early radiative simulations found rotational support exceeding random-motion support over $0.1$--$0.5\,R_{500c}$ in relaxed clusters \cite{fang09_icmrot,lau09_icmmotion}, whereas recent simulations find non-turbulent motions (without isolating out the rotation) contribute substantially to the non-thermal budget \cite{saha26_nth,lau26_nth}. A measurement of the gas rotation profile would constrain its contribution to the hydrostatic mass bias \cite{bartalesi24_coolcore}.
Cluster gas properties are also important for environmental processes that shape galaxy evolution, such as ram-pressure stripping, in which a galaxy’s cold gas is removed by the pressure of the surrounding ICM \cite{gunn72_env,boselli22_env}, which scales as the square of the galaxy’s velocity relative to the ICM \cite{gunn72_env,mccarthy08_ram}, so understanding the gas’ kinetic state is vital for understanding this picture.

Beyond its role in the kinetic state within clusters, halo angular momentum is also interesting as a statistical tracer of the initial conditions and of the cosmic web, as it correlates with clustering and environmental properties in mass-dependent ways.
First, at the high-mass end, higher-spin haloes exhibit tighter spatial clustering than lower-spin haloes, referred to as the halo spin bias, which increases with halo mass \cite{bett07_spinsim,gao07_spinbias,salcedo18_secondarybias,satopolito19_spinbias,johnson19_spinbias,tucci21_spinbias,montero21_spinbias}.
Second, mass-dependent correlations between spin and filament directions are also observed \cite{bailin05_spinfil,aragon07_spinfil,zhang09_spinfil,jones10_spinfil,pichon11_spinfil,codis12_spinfil,tempel13_spinfil,laigle15_spinfil,aragon14_spinfil,codis15_spinfil,kraljic21_alignment},  where lower (higher) mass haloes tend to have a spin parallel (perpendicular) to filaments with a redshift-dependent transition mass of $\sim 5\times 10^{12} M_\odot$ at $z=0$ \cite{codis12_spinfil,codis15_spinfil}; galaxy clusters therefore lie well within the perpendicular regime.
Third, correlations between the spin and shape of clusters are also observed in simulations, where the spin axis is preferentially perpendicular to the major axis of the cluster \cite{faltenbacher02_spinshape,porciani02_ttt,bailin05_spinfil,bett07_spinsim,fang09_icmrot,lau11_shape}.
Fourth, a detectable shape alignment of spiral galaxies could be used to reconstruct the initial tidal fields through their tidally-induced spin alignments \cite{heavens00_alignment,catelan01_alignment,lee00_alignment,pen00_alignment,crittenden01_alignment,lee11_alignment}. 
Observations of gas angular momentum in groups and clusters would therefore inform the studies of hydrostatic mass bias, and of clustering bias and spin alignments in the high-mass end, as well as lending evidence to the correct reconstruction of the local large-scale structure \cite{wang16_elucid,sheng25_elucidmanga,yang26_rksz}.

The rotational motion of cluster member galaxies or the ICM has been explored with multiple observables: 1.\ relative spectroscopic redshifts of member galaxies where early works have identified A2107 as a rotating cluster \cite{materne83_abell,oegerle92_abell,song18_abell}, 
and catalogs of rotating clusters are compiled \cite{hwang07_rotsdss,manolopoulou17_galrot,tang25_obs} from spectroscopic surveys e.g.\ SDSS and 2dFGRS based on criteria such as merger history, rotation velocity to velocity dispersion ratio, or number of member galaxies, 2.\ in the inner regions, centroid shifts and broadening of X-ray emission lines encode the rotational and turbulent motions of the ICM, and could potentially be measured by the XRISM/Resolve
X-ray spectrometer down to $\sim 10$ kms$^{-1}$ resolution \cite{bartalesi24_coolcore,bartalesi25_xcop,bartalesi26_galrot},
3.\ the bulk motion of free electrons in the ICM scatters off CMB photons and induces an intensity shift proportional to the line-of-sight integrated electron momentum, known as the kinetic Sunyaev-Zel'dovich (kSZ) effect; coherent ICM rotation induces a dipole signature in the observed intensity at microwave wavelengths \cite{cooray02_ksz,chluba02_ksz,altamura23_ksz,monllor24_szrot,baxter19_rksz,goldstein25_rksz,yang26_rksz}, which we henceforth call the rotational kSZ (rkSZ) effect and is the focus of this work.

The rkSZ signal was first computed analytically for clusters in Refs.\ \cite{cooray02_ksz,chluba02_ksz}.
Efforts towards detection rely on oriented stacking with respect to an external spin proxy, e.g.\ from integral-field spectroscopy \cite{bundy15_manga, zorrilla20_rksz}, and from member-galaxy dynamics of individual massive clusters \cite{manolopoulou17_galrot,baxter19_rksz, goldstein25_rksz}, or through constrained realization reconstruction of the local large-scale structure \cite{yang26_rksz}, while multipole expansion \cite{monllor24_szrot} and matched-filter estimators \cite{zorrilla20_rksz} have also been explored for this signal.
Given the small amplitude of the rkSZ signal (our model gives $\sim$ 6-7$\,\mu$K peak amplitude at $10^{15}\, M_\odot$ and $z=0.5$; see Fig.\ \ref{fig:1d_spatial}), existing observational studies \cite{baxter19_rksz,goldstein25_rksz,yang26_rksz} adopt an oriented stacking approach, where cluster cutouts are aligned by their rotation axis proxies and stacked to average down the anisotropies from the lensed CMB, instrumental noise and other extragalactic components that swamp the rkSZ dipole signal. 

In this work, we will study the prospects of measuring the rkSZ effect with upcoming data from Simons Observatory (SO). Conceptually, our formalism estimates a vector field tracing the projected cluster angular momentum field. To achieve this, we provide a simulation-validated model of cluster rotation and the amplitude and shape of the associated rkSZ signal, then we present a careful treatment of foregrounds (tSZ, kSZ, CIB, lensed CMB) and SO noise with our optimal multifrequency matched filter.
This paper is organized as follows:

Section~\ref{sec:modeling} introduces a minimal, simulation-motivated model of cluster rotation that depends only on the electron density profile and the gas spin parameter. 

Section~\ref{sec:rksz} turns it into the rkSZ signal (its real-space and Fourier profiles, peak amplitude and integrated statistics) and validates it against TNG-Cluster and TNG300-1 down to $10^{13.5}\,M_\odot$ at $z=0$, reproducing the integrated amplitude and its mass scaling.

Section~\ref{sec:extraction} derives the estimator. We construct a matched filter, linear in the temperature map, that recovers the per-cluster dipole vector and accounts for the scale-dependence of foreground and noise, together with a multifrequency form and a tensor variant for proxies that supply only the rotation axis.

Section~\ref{sec:forecast} forecasts the detectability for SO. Its central result is that the optimal foreground cleaning is mass dependent: the signal separates into an instrument-noise--dominated regime, which carries most of the cumulative SNR, and a thermal-SZ--dominated regime at the highest masses, where single-cluster SNR is the highest.
	 
Section~\ref{sec:discussion} discusses the modeling uncertainties across simulations, the foreground dipoles---from the lensed CMB and the moving-lens effect---that should be uncorrelated with the rkSZ dipole, and the detection strategies when targeting clusters of different mass ranges.

Section~\ref{sec:summary} summarizes our results and outlines applications of
the gas angular-momentum field as a cosmological observable.

Appendix~\ref{app:cov} derives the analytical covariance of the filtered data
for the vector and tensor estimators, including the clustered contribution of a
miscentered cluster tSZ profile and the effect of miscentering on the rkSZ
template. Appendix~\ref{app:filter_vec} derives the vector matched filter, its
homogeneous-noise limit and its multifrequency form.
Appendix~\ref{app:extract} describes how we extract the shell angular momenta,
tangential velocities and integrated rkSZ statistics from TNG-Cluster and
TNG300-1.

\section{Modeling cluster rotation}
\label{sec:modeling}

We construct a simple analytical model for cluster rotation as a function of mass and redshift. This model has four assumptions, stated and motivated in Sec.\ \ref{sec:assumptions}. The modeled quantities are shown in Figs.\ \ref{fig:beta} and \ref{fig:vtan_tng} in the halo mass range of $10^{14.25}$ to $10^{15.5}\, M_\odot$ and redshift $z=0$, 0.5, and 1.0, validated on TNG-Cluster \cite{nelson19_tng,nelson24_tngcluster}. In Sec.~\ref{sec:rksz} we use this rotation model to build the rkSZ
signal, and compare its integrand and integrated amplitude with simulated
haloes at $z=0$ (Figs.~4 and~7). For this comparison we add TNG300-1 haloes
with $M_{200c}\geq10^{13.5}\,M_\odot$ to the TNG-Cluster sample, which extends
the tested range down to $10^{13.5}\,M_\odot$. Our forecasts at lower masses
extrapolate the model (Sec.~\ref{sec:forecast}).
	   Appendix \ref{app:extract} discusses how we extract the relevant quantities from the simulation.

    TNG300-1 (roughly 300 comoving Mpc per side) is part of the IllustrisTNG suite of cosmological magnetohydrodynamical simulations of galaxy formation \cite{nelson18_tng,pillepich18a_tng,naiman18_tng,springel18_tng,marinacci18_tng}.
    Each simulation evolves gas, stars, dark matter and supermassive black holes from $z=127$ to $z = 0$ using a comprehensive galaxy formation model \cite{weinberger17_tng, pillepich18b_tng}. 
	TNG-Cluster is an addition to the IllustrisTNG suite, consisting of 352 zoom simulations of cluster regions ($\log_{10}(M_\text{200c}/M_\odot) \gtrsim$ 14.3–15.4 at $z = 0$) selected from a gravity-only 1 Gpc parent box by the halo mass at $z=0$, and re-simulated at roughly TNG300-1 resolution (baryon mass resolution of $1.2 \times 10^7 M_\odot$).
	While our simple model is consistent with TNG-Cluster+TNG300-1, we note in Sec.\ \ref{sec:discuss_sim_diff} that the radial angular velocity distribution is a modeling systematic that varies across simulations (as we compare with Baldi et al.\ \cite{baldi17_rot} and Altamura et al.\ \cite{altamura23_ksz}). 

\subsection{Definitions and Geometry}
\label{sec:definitions}

We adopt the following notation throughout.
Distances are physical; dimensionless radii are normalised by $R_{200c}$.
\begin{figure}
	\includegraphics[width=0.99\linewidth]{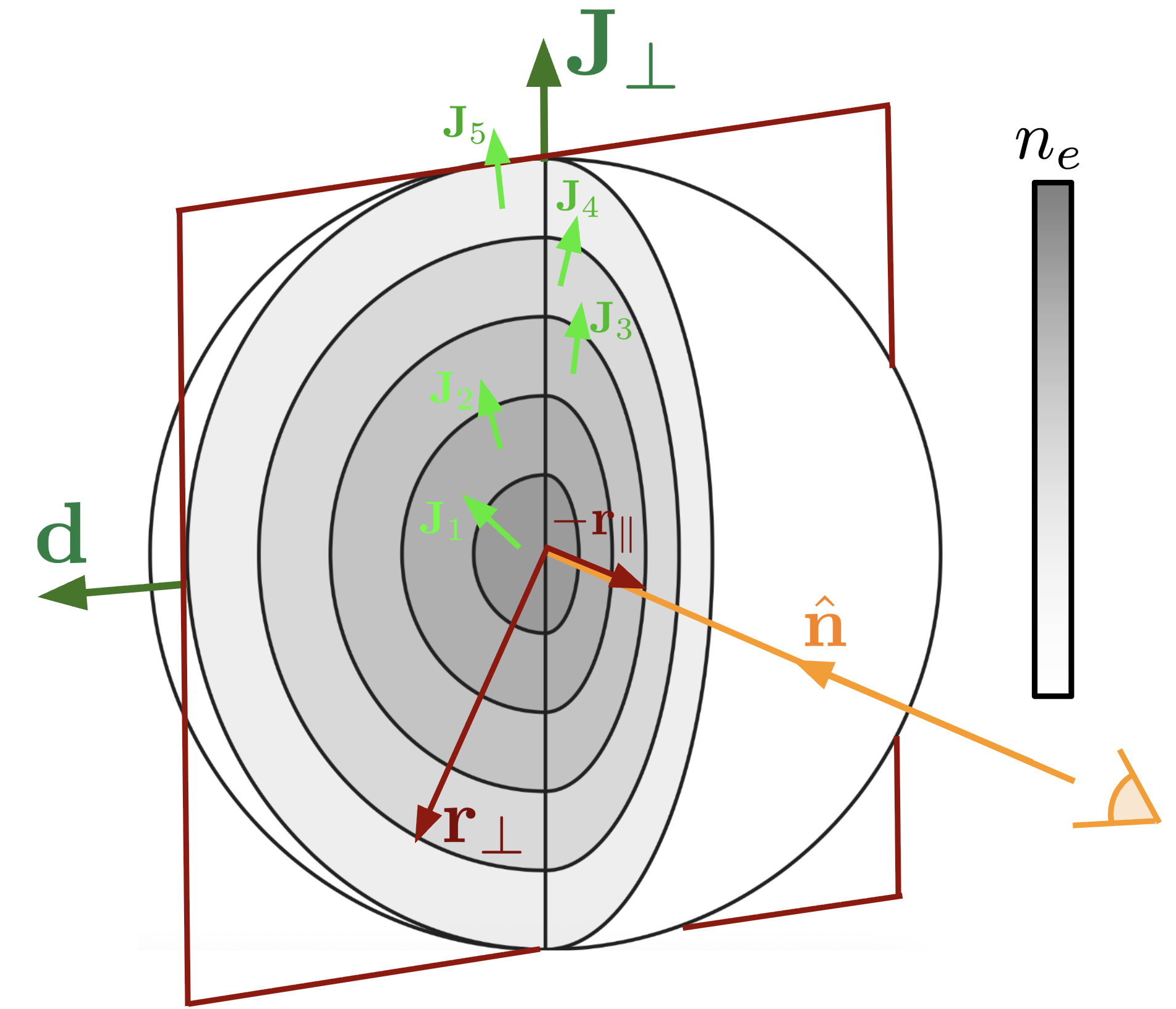}
	\caption{The ICM is modeled as concentric shells, labeled by $i$. Each shell is assumed to rotate as a solid body with a shell-dependent angular velocity $\omega_i$ and angular momentum $\hat{\mathbf{J}}_i$ (light-green arrows), and within a shell the electron distribution is assumed uniform with free-electron density $n_{e,i}$ (grayscale). The total angular momentum enclosed by $R_{200c}$ is  $\mathbf{J}$. Given a line-of-sight direction  $\hat{\mathbf{n}}$, the projected angular momentum is  $\hat{\mathbf{J}}_\perp$ and the rkSZ dipole direction is  $\mathbf{d}\equiv \mathbf{J}\times\hat{\mathbf{n}}$.
    Any point in the cluster can be identified with vectors (brown arrows) $\mathbf{r}_\parallel$ pointing along $\hat{\mathbf{n}}$, and $\mathbf{r}_\perp$ lying within the (brown) plane transverse to $\hat{\mathbf{n}}$.}
	\label{fig:schematic1}
\end{figure}

\begin{itemize}

  \item Radial, line-of-sight, and projected distances from the cluster centre are respectively
        $r$, $r_\parallel$, $r_\perp$, with dimensionless counterparts
        $x \equiv r/R_{200c}$,
        $x_\parallel \equiv r_\parallel/R_{200c}$,
        $x_\perp \equiv r_\perp/R_{200c}$. Their vector counterparts are boldfaced.
	The line-of-sight direction is $\hat{\mathbf{n}}$. See Fig.\ \ref{fig:schematic1} for a schematic diagram of the geometry.

\item The ICM is decomposed into concentric spherical shells with inner and outer radii $r_{i}-\Delta r/2$ and  $r_i+\Delta r/2$. In calculations, shell quantities are either denoted by discrete shell-label $i$, or as a continuous quantity as a function of  $r$ or  $x$. For example, the shell free-electron mass, free-electron number density, angular momentum, and angular velocity are respectively $M_{e,i}, n_{e,i}, \mathbf{J}_i, \omega_i$ in discrete labeling. See Fig.\ \ref{fig:schematic1} for a schematic diagram of the cluster model.

  \item The total free-electron angular momentum within $R_{200c}$ is $\mathbf{J}$ (without label), 
        with unit vector $\hat{\mathbf{J}}$.
        The rkSZ dipole points along
	${\mathbf{d}} \equiv {\mathbf{J}} \times \hat{\mathbf{n}}$. The projected angular momentum is $\mathbf{J}_\perp$. See Fig.\ \ref{fig:schematic1}.

  \item Shell-cluster spin misalignment angle factor $\beta_i$, describing the average deviation angle between the total free-electron angular momentum vector within $R_{200c}$, $\hat{\mathbf{J}}$, and that within each shell, $\hat{\mathbf{J}}_i$
\begin{align}
	\beta_i &= \hat{\mathbf{J}} \cdot \hat{\mathbf{J}}_i \; .
            \label{eq:beta_def}
\end{align}
We fit the distribution of $\beta_i$ as  $\bar\beta(x)$ in the next subsection.

  \item The Bullock spin parameter for the gas component is defined as
        \cite{bullock01_spinsim}:
        \begin{align}
		\lambda_\text{gas} \equiv
		\frac{|\mathbf{J}_\text{gas}|}{\sqrt{2}\,M_{\text{gas},<200c}\,R_{200c}\,V_{c,200c}} \;,
            \label{eq:bullock}
        \end{align}
	where $M_{\text{gas},<200c}$ denotes the mass of gas within $R_{200c}$, and $V_{c,200c} \equiv \sqrt{GM_{200c}/R_{200c}}$ is the circular velocity
	at $R_{200c}$. It follows a log-normal distribution
\begin{align}
	P(\lambda_\text{gas}) = \frac{1}{\sqrt{2\pi} \sigma_\lambda \lambda_{\text{gas}}} \exp\left(  - \frac{\ln^2(\lambda_\text{gas}/\lambda_{0,\text{gas}})}{2\sigma^2_\lambda} \right)\; ,
\end{align}
where we take $\lambda_{0,\text{gas}}=0.046$ and $\sigma_\lambda = 0.609$ based on Ref.\ \cite{gottloeber07_spinsim}. $\lambda_{0,\text{gas}}$ is also the median for the log-normal distribution.

\end{itemize}


\subsection{Modeling assumptions to the mean rkSZ signal}
\label{sec:assumptions}

\subsubsection*{Assumption 1: Per-shell solid-body rotation}

We model the ICM as a set of concentric spherical shells, each rotating
independently like a solid body about a direction $\hat{\mathbf{J}}_i$ that may
differ from the global spin axis $\hat{\mathbf{J}}$.
This is a \emph{per-shell} solid-body model: adjacent shells can rotate at different
speeds and about different axes.

To relate the shell angular momenta with the angular velocity, each shell $i$ is treated as a uniform-density solid body, giving
\begin{align}
    J_i = \frac{2}{3} M_{e,i} r_i^2 \omega_i \; .
    \label{eq:Ji_omegai}
\end{align}
Within each shell, the implicit assumption is that turbulent and random motions carry no net angular momentum.
The total projected angular momentum within $R_{200c}$ is
\begin{align}
    J \equiv \mathbf{J}\cdot\hat{\mathbf{J}}
    = \sum_{r_i \leq R_{200c}} \frac{2}{3} M_{e,i} r_i^2 \omega_i\beta_i \;.
    \label{eq:Jtot}
\end{align}

\subsubsection*{Assumption 2: Spin misalignment profile}
\begin{figure}
	\includegraphics[trim={0.0cm 0.0cm 0.0cm 0.7cm}, clip, width=\linewidth]{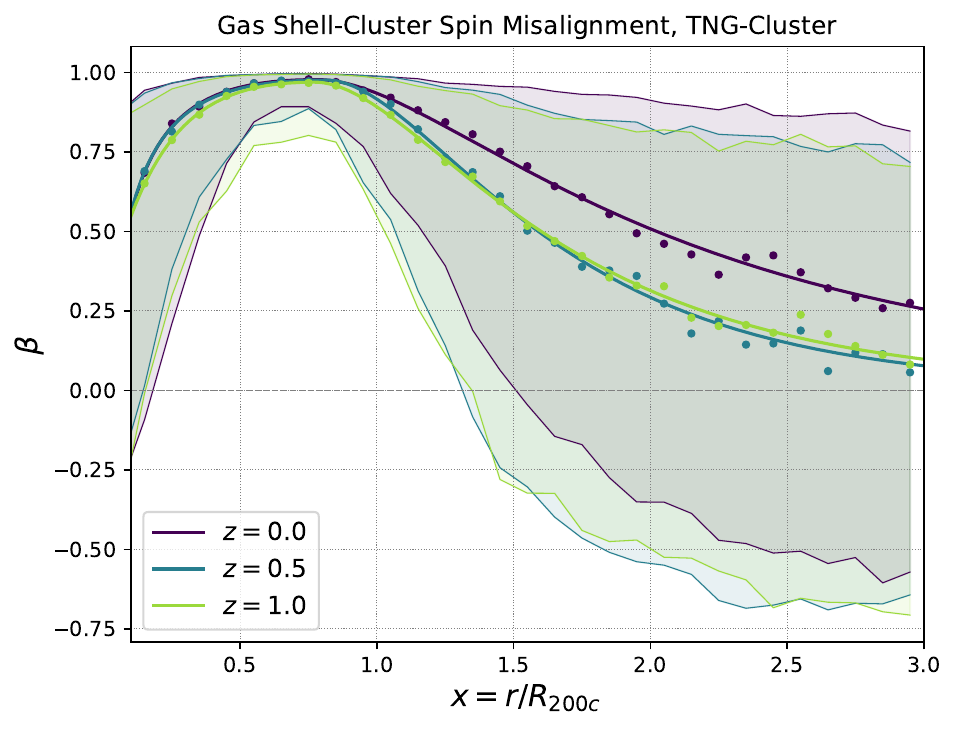}
	\caption{The best fit spin misalignment $\bar\beta(x)$ to the median of the distribution of $\beta_i$ (dots and shaded bands denote the median and the 16th to 84th percentile range respectively) from TNG-Cluster at three redshifts, $z=0$ (purple),  $z=0.5$ (blue), and $z=1.0$ (green). The piecewise functional form used for  $\bar\beta$ is described in Eq.\ (\ref{eq:beta_model}), with an inner and outer slope that can be independently varied. 
The amplitude of the angular velocity depends only weakly (a $\lesssim5\%$ effect on $v_\text{tan}/V_\text{circ,200c}$; compare the ranges quoted below Eq.\ \eqref{eq:avg_J}) on the shape of $\bar\beta$ at $x<1$.}
	\label{fig:beta}
\end{figure}

We compute $\beta_i$ for clusters in TNG-Cluster, and fit a profile  $\bar{\beta}(r)$ to the median of the distribution of  $\beta_i$. Motivated by the log-normal-like shape of the radial profile, we construct a piecewise profile from two log-normals, one controlling the inner (outer) slope with width $\sigma_\text{L}$ ($\sigma_\text{R}$):
\begin{align}
	\bar\beta(x) &=
	\begin{cases}
		(A-b) \exp\left[ - {\ln^2(x/x_0)}/{2\sigma^2_\text{L}} \right] + b \; , &\mbox{$x\leq x_0$} \\
		A \exp\left[ - {\ln^2(x/x_0)}/{2\sigma^2_\text{R}} \right]  \; , &\mbox{$x> x_0$}
	\end{cases}
    \label{eq:beta_model}
\end{align}
where the exterior profile ($\sigma_\text{R}$) describes the decreasing correlation of angular momenta in shells beyond the virial radius with the cluster angular momentum; the interior profile ($\sigma_\text{L}$, $b$) models the spin-misalignment within the cluster. The median position with greatest alignment is $x_0$ with amplitude $A$.
\begin{table}[ht]
	\setlength{\tabcolsep}{7pt}
\centering
\caption{Best-fit parameter values of Eq.\ (\ref{eq:beta_model}) to the median of the distribution of $\hat{\mathbf{J}}\cdot\hat{\mathbf{J}}_i$ from all TNG-Cluster clusters at three redshifts. We take the $z=0.5$ values as our fiducial $\bar\beta$ profile for all mass and redshifts.}
\begin{tabular*}{\linewidth}{@{\hspace{\tabcolsep}\extracolsep{\fill}} cccccc}
\toprule
$z$ & $A$ & $x_0$ & $\sigma_L$ & $\sigma_R$ & $b$ \\
\midrule
$0.0$ & $0.98$ & $0.78$ & $2.35$ & $0.82$ & $-0.35$ \\
$0.5$ & $0.98$ & $0.81$ & $2.57$ & $0.58$ & $-0.50$ \\
$1.0$ & $0.97$ & $0.77$ & $1.37$ & $0.63$ & $+0.34$ \\
\bottomrule
\end{tabular*}
\label{tab:beta}
\end{table}
The best-fit parameter values to the median is tabulated in Table \ref{tab:beta}, and Fig.\ \ref{fig:beta} shows the best fit $\bar\beta$ to TNG-Cluster data at three redshifts. Our analysis below takes the $z=0.5$ values as the profile for  $\bar\beta$ for all mass and redshifts, noting from Fig.\ \ref{fig:beta} that the inner profile is similar across redshifts, while the $z=0.5$ curve has a faster convergence to zero as $x\gg 1$, so that $\bar\beta$ acts as a taper to localize the subsequent rkSZ integrals radially.

\subsubsection*{Assumption 3: Modeling the electron number density $n_e$}
\label{sec:ne}

We model the free-electron number density $n_e(x)$ with the generalized-NFW (gNFW) fitting form of Battaglia \cite{battaglia16_ne}, evaluated at the cluster mass and redshift:
\begin{align}
	n_{e,\text{fit}} = \frac{f_b \rho_\text{crit}(z)}{\mu_e m_p} \rho_0 \left( \frac{x}{x_c} \right)^\gamma \left[1+\left( \frac{x}{x_c}  \right)^\alpha \right]^{-(\beta+\gamma)/\alpha} \nonumber
\end{align}
where $x\equiv r/R_{200c}$, $m_e$ is the electron mass, $\mu_e=2/(1+X_\mathrm{H})\approx1.14$ is the  mean molecular weight per free electron, $X_H\approx 0.76$ is the hydrogen fraction, $m_p$ the proton mass, $\rho_\text{crit} = 3H^2_0 \left[ \Omega_m(1+z)^3 + \Omega_\Lambda \right]/8\pi G$ is the critical density, $f_b\equiv \Omega_b/\Omega_m$ is the universal baryon fraction, $x_c=0.5$ and  $\gamma=-0.2$ is chosen to be fixed, and the fit parameters $\alpha,\beta,\rho_0$ have the following mass-redshift dependent form,
 \begin{align}
	 A = A_0\left( \frac{M}{10^{14}M_\odot} \right)^{\alpha_m} (1+z)^{\alpha_z} \; ,
\end{align}
where $A\in\{ \alpha,\beta,\rho_0 \}$. We use the parameters fitted for simulations with AGN feedback in Table 2 of Ref.\ \cite{battaglia16_ne}.
This is our fiducial choice throughout the filtering and forecast sections; deviations of the simulated $n_e$ from this fit contribute to the residuals in Figs.\ \ref{fig:nevtan_tng} and \ref{fig:Sx}.

We need not model the 2-halo contribution to $n_e$ since the spin-misalignment profile $\bar{\beta}$ acts as a taper with values $\sim 0.27$ ($\sim 0.1$) at $2R_{200c}$ ($3 R_{200c}$). In any case, the cluster outskirts $\gtrsim 2R_{200c}$ reaching out to the 2-halo regime contributes negligibly to our rkSZ modeled quantities (see Fig.\ \ref{fig:nevtan_tng}).

We note that recent kSZ measurements around group-scale haloes \cite{amodeo21_feedback,qu26_ksz} disfavor the Battaglia $n_e$
fits~\cite{battaglia16_ne}
and prefer stronger feedback than IllustrisTNG~\cite{amodeo21_feedback,hadzhiyska25_ksz,berni25_ksz}.
For this reason, although we compare our model with TNG-Cluster and TNG300-1
down to $10^{13.5}\,M_\odot$, our SNR forecasts use only haloes with
$M_{200c}\geq10^{14}\,M_\odot$ to minimize the baryonic-feedback uncertainty,
and results shown below this mass should be regarded as extrapolations. We
discuss the implications of baryonic feedback in Sec.~\ref{sec:discussion}.

\subsubsection*{Assumption 4: Radial profile of specific angular momentum}

We model the free electron specific angular momentum to scale linearly with radius,
\begin{align}
	J_i / M_{e,i} \propto r_i \; .
    \label{eq:alpha_1}
\end{align}
This is the radial distribution for dark matter haloes by Barnes \& Efstathiou \cite{barnes87_ttt}, and consistent with more recent studies on the dark matter component e.g.\ Refs.\ \cite{bullock01_spinsim,bett10_jdist, baldi17_rot}. We assume that the gas angular momentum follows the same scaling. 
With the solid-shell assumption above, this gives a \emph{constant} equatorial tangential speed\footnote{
Note our definition of $v_\text{tan}$ is $3/2$ times larger than the definition in Baldi et al.\ \cite{baldi17_rot}, which has the physical interpretation of a weighted angular momentum.
}:
\begin{align}
    v_{\rm tan} \equiv r_i\,\omega_i = \mathrm{const.}
    \label{eq:vtan_const}
\end{align}
Substituting this into the definition of the spin parameter allows us to relate $\lambda_e$ (which equals $\lambda_\text{gas}$), the circular velocity $V_\text{circ,200c}$, and the equatorial tangential speed  $v_\text{tan}$:
 \begin{align}
	 \lambda_\text{gas} &= \frac{ J_e}{\sqrt{2} M_{e,<R_{200c}}R_{200c}V_{c,200c}} \\
	   &\approx \frac{2}{3\sqrt{2}} \frac{v_\text{tan}}{V_\text{circ,200c}} \frac{4\pi \int_0^1 n_e x^3 \bar\beta dx }{4\pi \int_0^1 n_e x^2 dx} \; ,
							      \label{eq:avg_J}
\end{align}
which we used Eq.\ \eqref{eq:bullock} and Eq.\ \eqref{eq:Jtot}, and approximating $\beta(x)$ with the median best-fit profile  $\bar\beta(x)$, and $n_e$ is the number density of free electrons. 
\begin{figure}
    \includegraphics[trim={0.0cm 0.0cm 0.0cm 0.8cm}, clip,width=\linewidth]{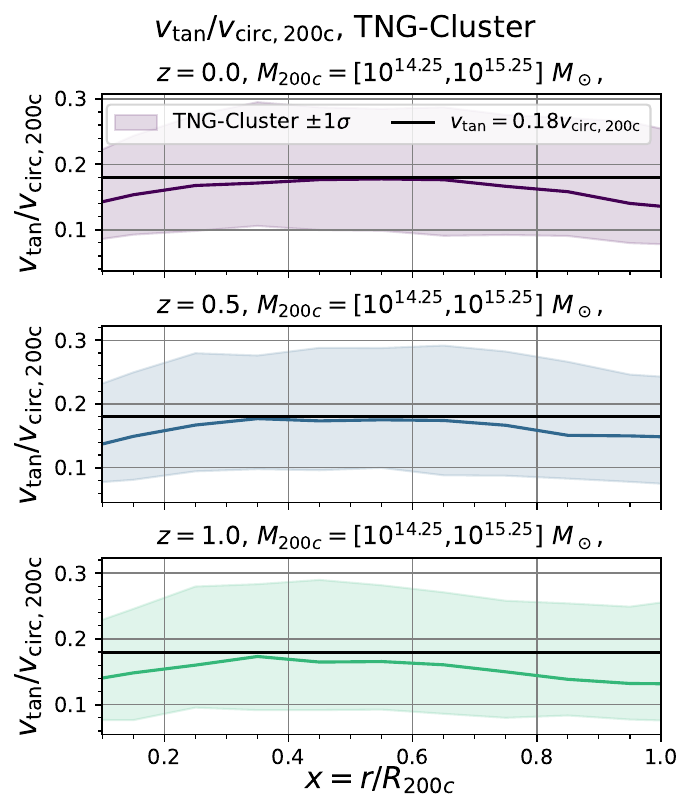}
    \caption{Modeled median equatorial tangential velocity $v_{{\rm tan}}=0.18 V_\text{circ,200c}$ (black), compared against the extracted $v_{\text{tan},i}$ using Eq.\ \eqref{eq:tng_extract} from TNG-Cluster at $z=0$ (top), $z=0.5$ (middle) and  $z=1$ (bottom) in the mass range $M_\text{200c}\in[10^{13.75},10^{15}] M_\odot$.
    The shaded region denotes the 16th-84th percentile across clusters, and the colored line denotes the median. While the simple estimate obtain an amplitude that agree with simulation for all considered mass and redshift, it cannot explain the radial features in the tangential velocity. In Fig.\ \ref{fig:fast_rotators} we compare with $v_\text{tan}$ profiles provided by Baldi et al.\ \cite{baldi17_rot} and Altamura et al.\ \cite{altamura23_ksz}.}
    \label{fig:vtan_tng}
\end{figure}

Adopting the Battaglia generalized NFW fit (our fiducial choice henceforth) for the electron number density profile, we can compute the ratio $v_\text{tan}/V_\text{circ,200c}$. For masses $M_{200c}\in[10^{13},10^{15}] M_\odot$ and redshifts  $z\in[0,1]$, the ratio falls in the range of  $v_\text{tan}/V_\text{circ,200c}\in[3.8\lambda_e,4.2\lambda_e]$. Because spin misalignment between shells allow more rotation for the same total angular momentum, setting $\bar\beta=1$ (no misalignment) reduces the ratio to $v_\text{tan}/V_\text{circ,200c}\in[3.7\lambda_e,4.0\lambda_e]$. As a zeroth order approximation, we take for our fiducial analysis
\begin{align}
	\frac{v_\text{tan}}{V_\text{circ,200c}} \approx 4\lambda_\text{gas} \; ,
\end{align}
that is, the ratio follows a log-normal distribution given by the spin parameter, substituting the median value of the spin-parameter as $\lambda_{\text{gas}}=0.046$ we have a median value of
\begin{align}
	\operatorname{Median}(v_\text{tan}) &=  0.18 V_\text{circ,200c} \; .
	\label{eq:vtan_to_vcirc}
\end{align}

Figure \ref{fig:vtan_tng} shows the comparison between our analytical estimate and results extracted from TNG-Cluster (App.\ \ref{app:extract}). 
Taking $j(r)\propto r$ can be seen as a simplifying assumption that allows $v_\text{tan}$ to be expressed as a constant over radius. Our constant ratio is in broad agreement with TNG-Cluster, and lies within the 16th-84th percentile for the redshift considered ($z=0,0.5,1$) in Fig.\ \ref{fig:vtan_tng}. A more detailed functional form would be required to describe the radial variation in the ratio that we also observe. 

We expect that the modeling of the gas angular velocity profile to contribute most of the systematic uncertainty of our study.
The profile changes significantly depending on whether one selects the full set of sample or a rotating subset \cite{baldi17_rot}.
In particular, Baldi et al.\ \cite{baldi17_rot} proposes a radial dependence
\begin{align}
	v_\text{tan}(r) &\propto \frac{r/r_0}{1+(r/r_0)^2} \; ,
	\label{eq:baldi_vtan}
\end{align}
for a subset of clusters with $\lambda_\text{gas}>0.07$. Altamura et al.\  \cite{altamura23_ksz} adopts similar functional form for their angular velocity.
Varying the angular velocity profile translates directly to a wide range of rkSZ profiles (see Appendix B of Ref.\ \cite{altamura23_ksz}.)
We compare predictions from different works with ours in Sec.\ \ref{sec:discuss_sim_diff}.

\section{Modeling the rotational kSZ signal}
\label{sec:rksz}
\begin{figure}
	\includegraphics[trim={0.0cm 0.0cm 0cm 0.9cm}, clip ,width=\linewidth]{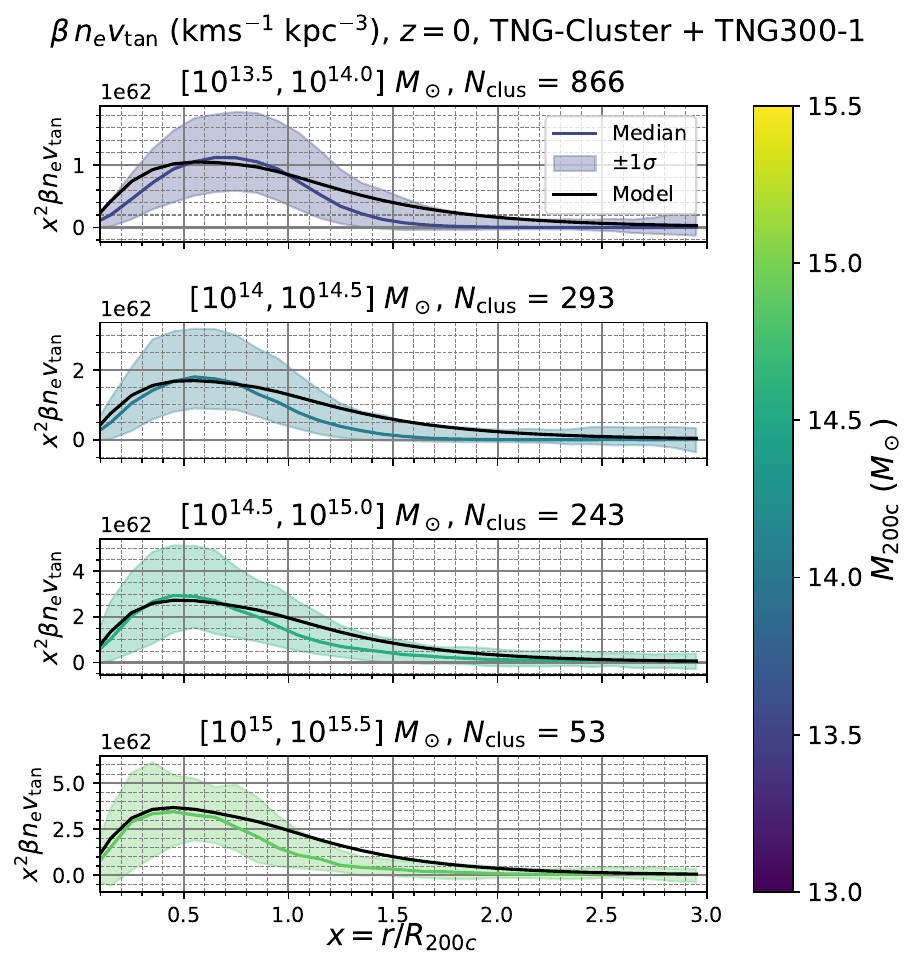}
	\caption{Integrals for the rkSZ effect (Eq.\ \ref{eq:Theta0}) involve the product of the electron-density-weighted tangential velocity profile $n_e v_{{\rm tan}} (x)$, and the shell misalignment factor $\bar\beta(x)$. 
	Our model is constructed as $\bar\beta n_ev_\text{tan} = 0.18 \bar\beta n_e V_\text{circ,200c}$ (with units of kms$^{-1}$\,kpc$^{-3}$), with  $\bar\beta$ in Eq.\ \eqref{eq:beta_model} (and fit values in Table \ref{tab:beta}), the 0.18 prefactor derived in \eqref{eq:vtan_to_vcirc}, and a Battaglia  $n_e$ profile. 
    The model (black line) at $z=0$ is compared against the same quantity extracted from TNG-Cluster+TNG300-1 $z=0$ haloes at shells $i$ of different radii, $n_{e,i}\hat{v}_{\text{tan},i}\hat{\mathbf{J}}\cdot\hat{\mathbf{J}}_i$, defined in App.\ \ref{app:extract}. The comparison is done in four $M_{200c}$-bins with bin centers  $\log (M_\text{200c}/M_\odot)=13.75,14.25,14.75,15.25$, where the shaded bands denote the 16th-84th percentiles, and the colored solid lines are the medians.
	The simple estimate lies within the 16th-84th percentile in all cases.}
    \label{fig:nevtan_tng}
\end{figure}
The (unbeamed) kSZ signal is proportional to its integrated column electron momentum projected along the line-of-sight direction:
\begin{align}
	\Theta(\mathbf{r}_\perp)
	&= -\frac{\sigma_T}{c} \int d r_\parallel n_e \boldsymbol{v}\cdot\hat{\mathbf{n}} \; ,
\end{align}
where $\boldsymbol{v}$ is the 3D velocity field of free electrons.
Let us model the kSZ induced by coherent rotation of electrons in each shell (as a function of $r$):
\begin{align}
	\Theta_\text{rkSZ}(\mathbf{r}_\perp)
	&= -\frac{\sigma_T}{c} \int \frac{dr_\parallel}{r} n_e v_\text{tan} \hat{\mathbf{J}}(r) \times \mathbf{r} \cdot \hat{\mathbf{n}} \\
	&= -\frac{\sigma_T}{c}\mathbf{r}_\perp\cdot\hat{\mathbf{n}}\times \int \frac{dr_\parallel}{r} n_e v_\text{tan}(r) \hat{\mathbf{J}}(r)  \; . 
\end{align}
We expect an average over clusters, with $\hat{\mathbf{J}}$ fixed, would average down the component of $\hat{\mathbf{J}}(r)$ orthogonal to 
$\hat{\mathbf{J}}$ due to shell misalignment. The median profile (at fixed mass and redshift) is defined as
\begin{align}
	\bar{\Theta}_\text{rkSZ}(\mathbf{r}_\perp) &= -\frac{\sigma_T}{c}\mathbf{r}_\perp\cdot\hat{\mathbf{n}}\times \hat{\mathbf{J}} \int \frac{dr_\parallel}{r}\bar\beta n_e v_\text{tan}(r)   \; . 
\end{align}
After some manipulation, it can be written in the form
\begin{align}
	\bar{\Theta}_\text{rkSZ}(\mathbf{r}_\perp) &= \hat{\mathbf{J}}\times\hat{\mathbf{n}}\cdot\hat{\mathbf{r}}_\perp R_{200c} \bar{\Theta}_0(x_\perp) \; ,
\end{align}
where $\bar{\Theta}_0$ is a circular-symmetric \emph{monopole} profile on the flat sky:
\begin{align}
	\bar{\Theta}_0(x_\perp) &\equiv \frac{2\sigma_T}{c} \int_{x_\perp}^{\infty} dx  \frac{x_\perp}{(x^2-x^2_\perp)^{1/2}} \bar\beta n_e v_\text{tan} \; . 
	\label{eq:Theta0}
\end{align}
The upper limit of infinity is not a concern, as  $\bar\beta(r)$ acts as a tapering function outside $x=1$.

\subsection{Fourier profile}

	We proceed by converting the rkSZ profile from physical coordinates $\mathbf{r}_\perp$ (in Mpc) to angular coordinates $\boldsymbol{\theta}$ on the flat sky, which relates by $\boldsymbol{\theta} = \mathbf{r}_\perp/D_A=\mathbf{x}_\perp R_{200c}/D_A$, where  $D_A$ is the angular diameter distance. We construct the usual Fourier pair $\boldsymbol{\theta}\leftrightarrow\Bell$ on the flat sky. 

	The rkSZ signal in Fourier space has the simple functional form below, determined by an amplitude $\mathcal{S}$ (discussed in detail in Sec.\ \ref{sec:integrated_statistic}), a dipolar structure ($-i\hat{\mathbf{d}}\cdot\hat{\Bell}$, where $\hat{\mathbf{d}}$ is the dipole direction), a geometric factor $|\hat{\mathbf{J}}\times\hat{\mathbf{n}}|$ dependent on the 3D orientation of the electron angular momentum, and a shape $F(\ell)$:
	\begin{align}
		\bar{\Theta}_\text{rkSZ}(\Bell) 
						&\equiv-i  |\hat{\mathbf{J}}\times\hat{\mathbf{n}}|\mathcal{S}   \hat{\mathbf{d}}\cdot \hat{\Bell} F(\ell) \; .
						\label{eq:rksz_fourier_unbeamed}
\end{align}
\begin{figure}
	\includegraphics[trim={0.0cm 0.0cm 0cm 0.9cm}, clip ,width=0.95\linewidth]{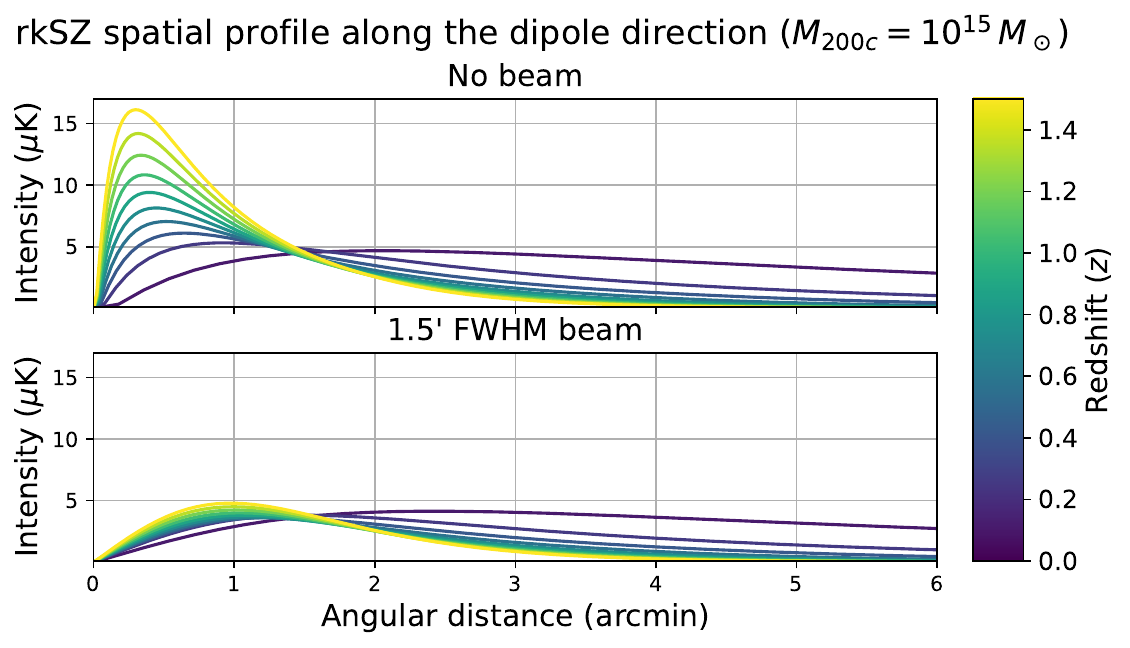}
\caption{At a fixed halo mass ($M_{200c} = 10^{15} M_\odot$), the modeled rkSZ spatial profile along the dipole direction is plotted for a range of redshifts (colour) for our fiducial model (see Fig.\ \ref{fig:nevtan_tng} for a description). The unbeamed profiles (top) and the 1.5' FWHM beam-convolved profile (bottom) are plotted. At this high mass, the peak unbeamed rkSZ amplitude is above 3 $\mu$K, and could exceed 10 $\mu$K at high redshifts.}
	\label{fig:1d_spatial}
	\includegraphics[trim={0.0cm 0.0cm 0cm 0.9cm}, clip ,width=0.99\linewidth]{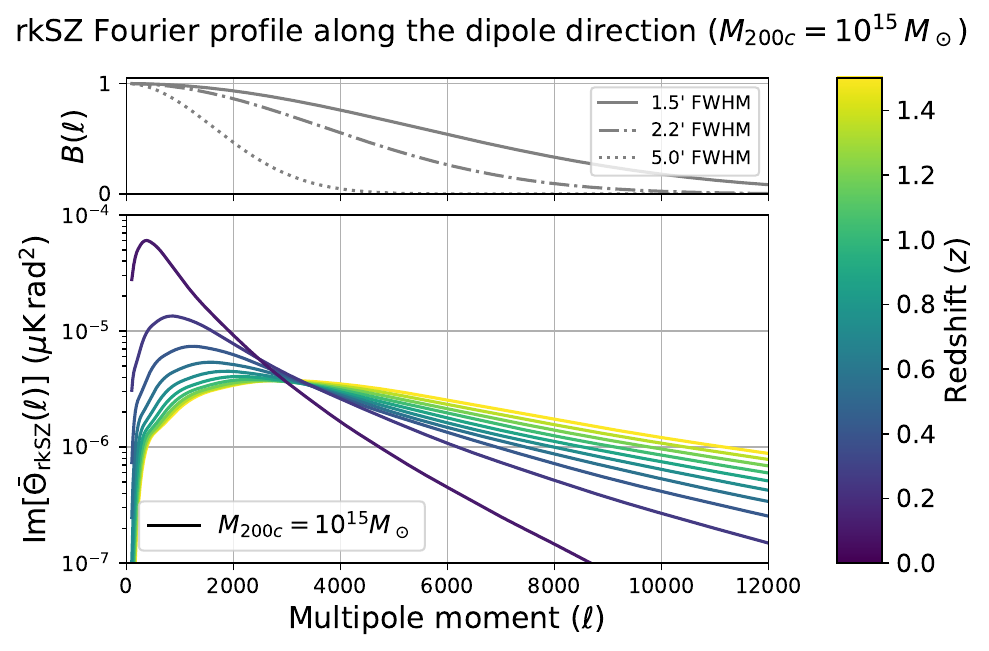}
	\caption{The corresponding Fourier profile $|\operatorname{Im}\bar\Theta_\text{rkSZ}(\Bell)|$ is plotted along the dipole direction (Eq.\ \eqref{eq:rksz_fourier_unbeamed}), together with representative Gaussian beams $\mathcal{B}(\ell)$ (top panel); the signal retains support out to $\ell\gtrsim8000$, especially at high redshift, suggesting more statistical power could be extracted with a higher-resolution beam.}
	\label{fig:1d_rksz}
\end{figure}
The beamed Fourier profile with azimuthally-symmetric beam $\mathcal{B}$ is
\begin{align}
	\bar{\Theta}^{\mathrm{beamed}}_{\mathrm{rkSZ}}(\Bell) 
						&=-i  |\hat{\mathbf{J}}\times\hat{\mathbf{n}}|\mathcal{S} \hat{\mathbf{d}}\cdot \hat{\Bell}\mathcal{B}(\ell) F(\ell) \; .
						\label{eq:rksz_fourier}
\end{align}
The dimensionless spectral shape $F(\ell)$ is derived as
 \begin{align}
	 F(\ell) &\equiv \frac{4\pi \sigma_T}{c} \frac{R^3_{200c}}{D^2_A\mathcal{S}} \int_0^\infty x^2 dx \bar\beta n_e v_\text{tan} j_1(R_{200c}\ell x/D_A) \; ,
\end{align}
whereas the amplitude $\mathcal{S}$ we will choose as the integrated rkSZ signal over an aperture size of $\theta=R_{200c}/D_A$ (to be defined in the next subsection):
 \begin{align}
	 \mathcal{S}
				    &\equiv \frac{\pi^2 \sigma_T}{2c} \frac{R^3_{200c}}{D^2_A} \int_0^{1} x^2dx \bar\beta n_e v_\text{tan} \; .
				    \label{eq:smodel}
\end{align}
Figure~\ref{fig:1d_rksz} shows the Fourier profiles along the dipole axis for $M_{200c}=10^{15}\,M_\odot$ at different redshifts, where the spectrum peaks at $\ell\approx500$ for $z\lesssim0.1$ and extends support out to $\ell\gtrsim 8000$ at higher redshifts.

\subsection{Peak rkSZ amplitude}
The maximum intensity, or the amplitude, of the rkSZ dipole is often reported to describe the strength of the signal. It is straightforward to calculate the un-beamed and beamed profile along the dipole direction $\hat{\mathbf{d}}$ where the maximum lies.

Including a symmetric instrumental beam $\mathcal{B}(\ell)$ and integrating
Eq.~\eqref{eq:rksz_fourier} over azimuthal angle to obtain the 1D profile along
the dipole axis $\hat{\mathbf{d}}$:
The beamed 1D profile is
\begin{align}
	\bar{\Theta}_\text{rkSZ}^\text{1D,beamed}(\theta_x\hat{\mathbf{d}})
						     &= 2\pi  |\hat{\mathbf{J}}\times\hat{\mathbf{n}}|\mathcal{S} \int \frac{\ell d\ell}{(2\pi)^2} J_1(\ell\theta_x) \mathcal{B}(\ell) F(\ell) \; .
    \label{eq:1D_beamed}
\end{align}
The peak of Eq.~\eqref{eq:1D_beamed} is the maximum observable rkSZ dipole
amplitude for a given instrument (see Fig.\ \ref{fig:1d_spatial}).

\subsection{Integrated rkSZ statistic}
\label{sec:integrated_statistic}
\begin{figure}
	\includegraphics[trim={0.0cm 0.0cm 0cm 0.9cm}, clip ,width=\linewidth]{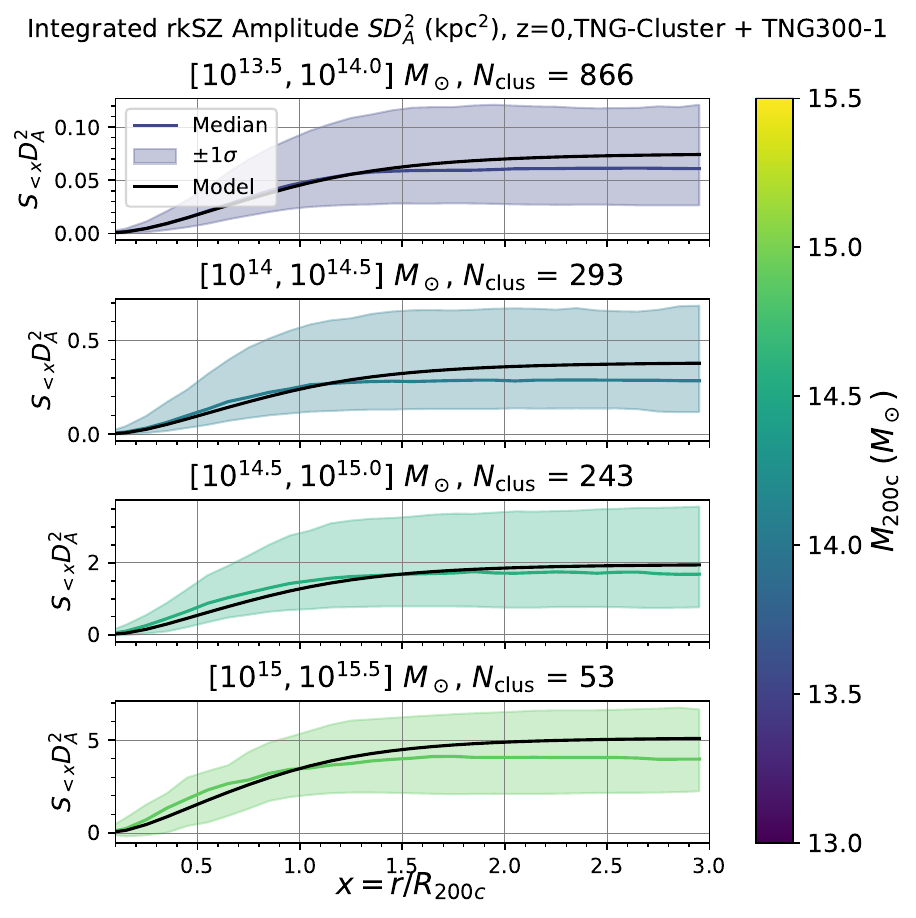}
	\caption{Comparison between the modeled integrated statistic $\mathcal{S}_{<x}^\text{model} D_A^2 = \pi^2 \sigma_T R^3_{200c}/2c \int_0^x (x')^2 dx' \bar\beta n_e v_\text{tan}$ (unit of kpc$^2$, Eq.\ \eqref{eq:smodel_x}) of the integrated rkSZ dipole contributed by shells of radii $<x$, against the extracted $\hat{\mathcal{S}}^\text{model,avg}_{<x}$ from TNG-Cluster and TNG300-1 $z=0$ haloes (see App.\ \ref{app:extract}). This is an integral of $x^2 \bar\beta n_e v_\text{tan}$ in Fig.\ \ref{fig:nevtan_tng}. Similar to Fig.\ \ref{fig:nevtan_tng}, we split the data into $M_{200c}$-bins, and the distribution 16th-84th percentiles and medians are plotted as colored shaded bands and solid lines respectively. The discrepancy in the lowest-mass bin (top panel) originates from that in the integrand, see Fig.\ \ref{fig:nevtan_tng}. The agreement between our model and the simulation-extracted integrated statistic within $R_{200c}$ is excellent. At $R_{200c}$, which defines $\mathcal{S}$ (Eq.~\eqref{eq:smodel}), model and median agree to within $\approx10\%$ in all mass bins}
    \label{fig:Sx}
\end{figure}
We can integrate the averaged rkSZ signal spatially with a cosine pixel weighting within an aperture $\mathcal{A}$ of radius $R$, oriented along the dipole direction  $\hat{\mathbf{d}}$. This is a cylindrical volume integral, denoted $\mathcal{S}^\text{cylin}_{<R}$. We take $\hat{\mathbf{J}}$ to be perpendicular to the line of sight when we evaluate the integrated statistics, $|\hat{\mathbf{J}}\times\hat{\mathbf{n}}|=1$; if spin directions are oriented isotropically we have  $\langle |\hat{\mathbf{J}}\times\hat{\mathbf{n}}|  \rangle = \pi /4 $, which we will factor into the SNR forecast below.
\begin{align}
	\mathcal{S}^\text{cylin}_{<R} &\equiv\hat{\mathbf{d}}\cdot\int_{\mathcal{A}}\hat{\boldsymbol{\theta}}   d^2\boldsymbol{\theta} \bar{\Theta}_\mathrm{rkSZ}(\boldsymbol{\theta}) \\ 
				    &= \pi \frac{R^3_{200c}}{D^2_A} \int_0^{R/R_{200c}} x_\perp dx_\perp \bar{\Theta}_0(x_\perp) \; .
\end{align}
With simulations, we can access instead a spherical volume integral of radius $R$ from the cluster center (note the integration limits), capturing only the dipole contribution from within the cluster:
 \begin{align}
	 \mathcal{S}^\text{model}_{<R}
				       &= \frac{2\pi\sigma_T}{c} \frac{R^3_{200c}}{D^2_A} \int_0^{R/R_{200c}} dx_\perp \int_{x_\perp}^{R/R_{200c}} dx \nonumber \\
				    &\phantom{ASDADSASDS}\times\frac{x_\perp^2}{(x^2-x^2_\perp)^{1/2}} \bar\beta n_e v_\text{tan} \\
				    &= \frac{\pi^2 \sigma_T}{2c} \frac{R^3_{200c}}{D^2_A} \int_0^{R/R_{200c}} x^2dx \bar\beta n_e v_\text{tan} \; .
				    \label{eq:smodel_x}
\end{align}
We take for the amplitude of the Fourier rkSZ profile in Eq.\ \eqref{eq:rksz_fourier} as
\begin{align}
	\mathcal{S}\equiv \mathcal{S}^\text{model}_{<R_{200c}} \; .
    \label{eq:S_Smodel}
\end{align}
Therefore,
\begin{align}
	F(\ell) &= \frac{8}{\pi} \frac{\int_0^\infty x^2 dx \bar\beta n_e v_\text{tan}j_1(R_{200c}\ell x/D_A)}{\int_0^1 x^2dx \beta n_e v_\text{tan}} \; .
	\label{eq:Fell}
\end{align}
Figure~\ref{fig:Sx} compares $\mathcal{S}^{\rm model}_{<x}$ with the same statistic extracted from TNG-Cluster and TNG300-1 at $z=0$ (see App.~\ref{app:extract} for extraction). At $R_{200c}$, the model $\mathcal{S}$ (Eq.\ \eqref{eq:S_Smodel}) is within $\approx10\%$ of the simulated median in every mass bin.

We can also extract the aperture filter statistic used in both observational and simulation work (e.g.\ Refs.\ \cite{baxter19_rksz,zorrilla20_rksz,montero21_spinbias,goldstein25_rksz,yang26_rksz}), where the sum of pixel intensities at one side of the projected rotation axis is subtracted from that of the other:
\begin{align}
	\mathcal{D}^\text{cylin}_{<R} &\equiv\int_{\mathcal{A}} \mathrm{Sign}[\hat{\mathbf{d}}\cdot \hat{\boldsymbol{\theta}}] d^2\boldsymbol{\theta}  \bar{\Theta}_\mathrm{rkSZ}(\boldsymbol{\theta}) \\ 
				    &= \frac{R^3_{200c}}{D^2_A} \int d\phi \frac{(\hat{\mathbf{d}}\cdot\hat{\boldsymbol{\theta}})^2}{|\hat{\mathbf{d}}\cdot\hat{\boldsymbol{\theta}}|} \int x_\perp dx_\perp \Theta_0(x_\perp) \\ 
				    &= \frac{4}{\pi} \mathcal{S}^\text{cylin}_{<R} \; .
\end{align}
where Sign$[\cdot]$ outputs $+1$ ($-1$) if the argument is positive (negative); this operator flips the sign of the rkSZ signal at one side of the rotational axis. We also have
\begin{align}
	\mathcal{D}^\text{model}_{<R} &= \frac{4}{\pi}\mathcal{S}^\text{model}_{<R} \; .
\end{align}
The factor $4/\pi$ arises from the azimuthal integrals
$\int d\phi |\cos\phi| = 4$ (for $\mathcal{D}$) and
$\int d\phi \cos^2\phi = \pi$ (for $\mathcal{S}$).

We can then compare the integrated $\mathcal{S}$-statistic we extracted from TNG-Cluster with that extracted by Montero-Dorta et al.\ \cite{montero21_spinbias} for the IllustrisTNG simulations for $\sim 5\times 10^4$ TNG300-1 haloes with $M_\text{200c} > 10^{11} M_\odot/h$ at $z=0$. 
	Their integrated statistic (Eq. 9) is the difference of the summed pixel intensity $\Delta T$ (units of $\mu$K) between two halves of a square map divided by a partition (labeled by $+$ and  $-$),  weighted by the pixel solid angle $A_\text{pixel}/D^2_A$, with the line of sight chosen such that $|\hat{\mathbf{J}}\times\hat{\mathbf{n}}|=1$, and $D_A$ artifically chosen at $z = 0.05$ for every halo (such that $D_A$ is not zero):
	\begin{align}
		S_\text{kSZ}^\text{Montero-Dorta} &= \frac{A_\text{pixel}}{D_A^2}\left[ \sum_{+} \Delta T - \sum_- \Delta T\right] \; ,
	\end{align}
	The statistic is therefore similar to our $\mathcal{D}^\text{model}_{<R}$-statistic, with the difference that they integrate over a square column where the face has sides $2L_\text{max}\times 2L_\text{max}$, where  $L_\text{max}$ is the maximum projected separation of any FoF gas cell from the halo centre of mass; instead of partitioning using projected rotation axis, the largest $S_\text{kSZ}^\text{Montero-Dorta}$ is selected out of 50 different orientations of the partition, after masking pixels more than $5\sigma$ from the mean of their half.
	The comparison with our $\mathcal{S}^\text{model}$ is done in Fig. \ref{fig:montero_dorta}.
    In the overlapping mass range of $5\times 10^{12}\,M_\odot\lesssim M_{200c}\lesssim 2\times10^{14}\,M_\odot$ which we extract our statistics, we obtain comparable mass scalings, while the factor of $\sim 2$ difference is potentially due to physical differences in azimuthal weighting ($\mathcal{S}$ vs $\mathcal{D}$), or geometry (integrating a spherical volume vs square column), or aperture size ($R_{200c}$ vs $L_\text{max}$ above), or the masking of pixel outliers in Montero-Dorta et al.

\subsection{Mass-redshift scaling for the integrated rkSZ statistics}

From Eq.~\eqref{eq:smodel}, using $V_{c,200c}\propto M_{200c}^{1/3}H(z)^{1/3}$, $v_{\rm tan}\sim V_{c,200c}$ (neglecting the residual mass and redshift dependence of $v_\text{tan}/V_{c,200c}$ visible in Fig.\ \ref{fig:vtan_tng})
(Eq.~\ref{eq:vtan_to_vcirc}), and
$\int_0^{R_{200c}} n_e\,r^2\,dr \propto M_{200c}\,f_b/m_e$,
we can derive a simple mass-redshift scaling for $\mathcal{S}$: 
\begin{align}
	\mathcal{S} D^2_A &\sim V_{c,200c} \int^{R_{200c}}_0 r^2dr n_e \\
			  &\sim V_{c,200c} M_{200c} f_b \\
			  &\sim M^{4/3}_{200c} H(z)^{1/3} f_b \; .
			  \label{eq:Smodel_scaling}
\end{align}
where $f_b$ is the baryon fraction. 
This $M^{4/3}$ scaling reflects the product of the circular velocity
($\propto M^{1/3}$) with the total free electron mass ($\propto M$) and
is consistent with the scaling extracted in Eq.\ \eqref{eq:sda_scaling}.

\begin{figure}
	\includegraphics[trim={0.0cm 0.0cm 0cm 0.68cm}, clip ,width=\linewidth]{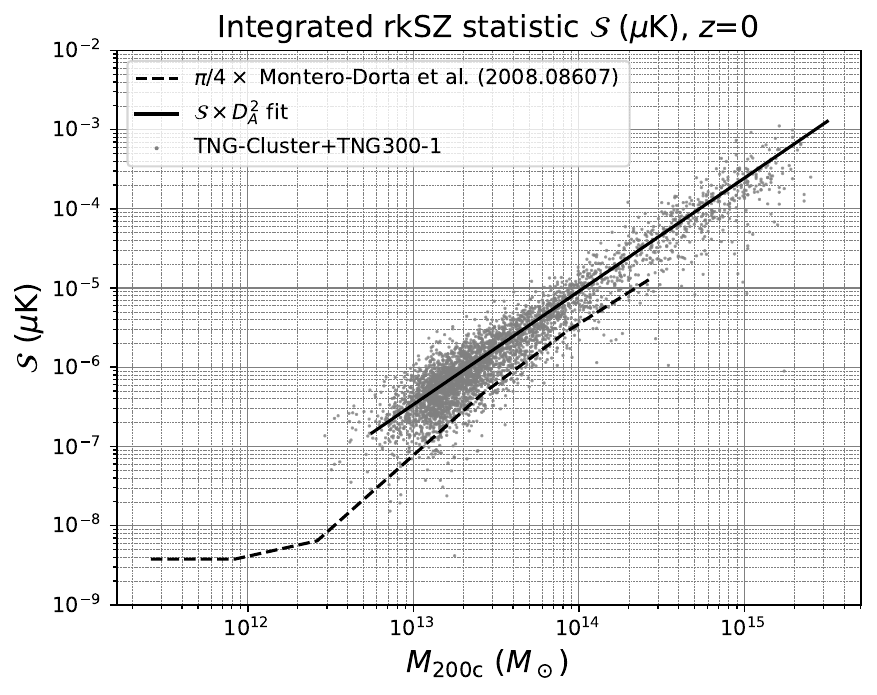}
	\caption{
		The statistic $\mathcal{S}D_A^2$ (Eq.\ \eqref{eq:smodel}) extracted for clusters in the $z=0$ snapshot of TNG-Cluster and the most massive 3500 TNG300-1 clusters (dots), as well as the $\mathcal{S}D_A^2$ scaling we obtained in Eq.\ \eqref{eq:sda_scaling} (dashed line), divided by $D_A^2(z=0.05)=208$ Mpc following Montero-Dorta et al.\ \cite{montero21_spinbias}.
		We multiplied the Montero-Dorta et al.\ statistic $S_\text{kSZ}^\text{Montero-Dorta}$ (dashed line, Eq.\ 9, Ref.\ \cite{montero21_spinbias}) by $\pi/4$ to convert it from a $\mathcal{D}$-statistic to a $\mathcal{S}$-statistic (see main text),
		The simulation lines of sight are chosen orthogonal to $\hat{\mathbf{J}}$, as done also in Montero-Dorta et al., so the extracted points carry $|\hat{\mathbf{J}}\times\hat{\mathbf{n}}|=1$.
		There are real differences in the statistics we extract: 1.\ we integrate a spherical volume of physical distance $R_{200c}$ to obtain  $\mathcal{S}$, while Montero-Dorta et al.\ integrate a rectangular column, and masks out outlier pixels  (see main text).
	}
	\label{fig:montero_dorta}
\end{figure}

The scaling of the $\mathcal{S}$-model constructed from $\bar\beta$,  $v_\text{tan}$, and  $n_e$ is in fact slightly steeper
(Fig.\ \ref{fig:montero_dorta}),
\begin{align}
	\log_{10}(\mathcal{S}D^2_A / {\text{Mpc}^2}) &= 1.43 \log_{10}(M_{200c}/M_\odot) - 26.86 \; .
	\label{eq:sda_scaling}
\end{align}

\section{Extracting the rkSZ dipole field from CMB-only observations}
\label{sec:extraction}
In the previous section, we have modeled the median rkSZ signal as a function of mass and redshift. In this section we discuss how to extract these dipole vectors optimally through a matched filter.
We also introduce a conceptually new angle of viewing the data analysis process: of the rotational kSZ dipole vector (or the projected angular momentum) field as an observable that can be extracted from CMB data alone. We can construct a dipole vector $\mathbf{V}(\hat{\mathbf{n}})$ or a headless dipole axis  $\mathbb{Q}(\hat{\mathbf{n}})$ to create a vector or tensor field. Practically, these are noisy fields where signal is extracted from cross-correlating respectively with spin-1 (e.g.\ rotational dipole fields) or spin-2 (e.g.\ tidal, shape fields). In a futuristic experiment one could envision using the angular distribution of this field to study the spin bias, spin-spin alignments, and global statistics of angular momentum. These are all rich and largely unexplored SNR-limited observational studies. The same formalism can be applied to other cosmologically-interesting dipoles such as the moving lens effect and the lensing dipole.

	We adopt the noise model of Simons Observatory \cite{so19_whitepaper}, using the 27, 39, 93, 145, 225, and 280 GHz noise channels of the Large Aperture Telescope (LAT). The noise spectra is modeled as a white-noise component, and an atmospheric part which scales as $N_\text{red}(\ell/\ell_\text{knee})^{\alpha_\text{knee}}$, where $\ell_\text{knee}=1000$ and $\alpha_\text{knee}=-3.5$. The channel pairs (27,39), (93,145), and (225,280) have highly correlated atmospheric noise with correlation coefficient assumed to be 0.9.
	The beam FWHM, atmospheric noise ampltide and the goal white-noise level are tabulated in Table \ref{tab:instrument_specs}. We assume a 5-year survey time with 20\% of the total observing time used for science analysis and sky fraction $f_\text{sky}=0.4$. 
	We compute the multifrequency noise spectra using the publicly-available \texttt{so\_noise\_models} package (v3.1.2)\footnote{\href{https://github.com/simonsobs/so_noise_models}{\texttt{https://github.com/simonsobs/so\_noise\_models}}}.

Extragalactic component maps are taken from the WebSky simulation \cite{stein20_websky}, a full-sky realization covering $0<z<4.6$ in a Planck 2018 cosmology providing correlated lensed CMB, tSZ, CIB and kSZ maps, where
	halos masses and positions are obtained by the mass-Peak Patch algorithm \cite{stein19_peakpatch}, the tSZ and kSZ maps are generated by painting hydrosimulation-calibrated profiles \cite{battaglia12,battaglia16_ne} onto halos, the lensing convergence by integrating the matter overdensity along
the line of sight, and
the CIB by populating halos and subhalos with central and satellite galaxies which emit as modified blackbodies, following a halo
model \cite{shang12_cib} calibrated to Herschel CIB power spectrum
measurements \cite{viero13_hermes}.
We take the 143 GHz CIB map from WebSky and scale its amplitude using an assumed spectral energy distribution to simulate CIB maps from other
frequency channels. 

\begin{table}[ht]
    \centering
    \caption{Simons Observatory LAT instrumental noise model taken from Table 1 and 2 of Ref.\ \cite{so19_whitepaper}. The white noise is taken to be the ``goal'' noise level.}
    \label{tab:instrument_specs}
    \begin{tabular}{lcccccc}
        \toprule
        Frequency Channel (GHz) & 27 & 39 & 93 & 145 & 225 & 280 \\
        \midrule
        FWHM (arcmin)       & 7.4 & 5.1 & 2.2 & 1.4 & 1.0 & 0.9 \\
	White noise ($\mu$K\,arcmin) & 52 & 27 & 5.8 & 6.3 & 15 & 37 \\
	$N_\text{red}$  ($\mu$K$^2$\,s) & 100 & 39 & 230 & 1,500 & 17,000 & 31,000 \\
        \bottomrule
    \end{tabular}
 
\end{table}

\subsection{Recovering the rkSZ dipole direction: the vector matched filter formalism}
Observational noise, including instrumental noise and foregrounds, exhibits strong dependence on scale. For example the primary CMB dominates at large scales, and instrumental noise at small scales. We therefore introduce a matched-filtered statistic where noisy scales are optimally down-weighted. Similar ideas of applying matched filtering and multipole expansion to rkSZ has been explored in Refs.\ \cite{zorrilla20_rksz, monllor24_rksz}.

The starting point of this section is the signal model of Eq.\ \eqref{eq:rksz_fourier_unbeamed},
$\bar{\Theta}_\text{rkSZ}(\Bell) =  -i  |\hat{\mathbf{J}}\times\hat{\mathbf{n}}|  \mathcal{S}\hat{\mathbf{d}} \cdot\hat{\Bell} F(\ell)$,
which is purely imaginary in Fourier space as the spatial profile is real-valued and odd. We write the (single-frequency) data $m(\Bell)$ as 
\begin{align}
	m(\Bell) &= -i  |\hat{\mathbf{J}}\times\hat{\mathbf{n}}|   \mathcal{S} \hat{\mathbf{d}} \cdot\hat{\Bell} F(\ell) + n(\Bell) \; , 
\end{align}
where recall $\hat{\mathbf{d}}\equiv \hat{\mathbf{J}}\times\hat{\mathbf{n}}/|\hat{\mathbf{J}}\times\hat{\mathbf{n}}|$ is the dipole direction (from negative to positive), the rkSZ  model amplitude is given in Eq.\ \eqref{eq:smodel}, and the expectation of the projection factor is $\langle | \hat{\mathbf{J}}\times\hat{\mathbf{n}} | \rangle=\pi/4$.
One should note that the filtering would no longer be optimal if the shape $F(\ell)$ of the mean rkSZ profile is mismodeled. For example, if a sample cut is applied to the catalog where only fast-rotators are selected, in which case the angular velocity follows a different radial dependence e.g.\ in Baldi \cite{baldi17_rot} and the alternative modeling choice should be substituted into  $F(\ell)$. Mismodeling of the signal amplitude affects the optimal weighting of objects to obtain the combined SNR.

The data analysis methodology can be interpreted in three steps:

\textbf{Step 1:} From data cutouts, group the Fourier modes $\Bell$ into $\bar\ell$-bins with $|\Bell|\in [\bar\ell-\Delta\ell/2,\bar\ell+\Delta\ell/2]$. In each $\bar\ell$-bin, perform an azimuthal weighting to the imaginary part of the Fourier modes (where the signal lies) by the radial vector  $\hat{\Bell}$. This procedure constructs a data vector $\mathbf{D}_{\bar\ell}$:
		\begin{align}
			\mathbf{D}_{\bar{\ell}}
						&\equiv \int_{\bar{\ell}\text{-bin}} \frac{d^2\ell}{(2\pi)^2} \operatorname{Im}[m(\Bell)]\hat{\Bell} \; , 
		\end{align}
        where the integral over $\bar\ell$-bin is shorthand for
        \begin{align}
            \int_{\bar\ell\text{-bin}}\frac{d^2\ell}{(2\pi)^2}\;[\,\cdot\,] \equiv \int^{\bar\ell+\Delta \ell/2}_{\bar\ell - \Delta\ell/2} \frac{\ell d\ell}{(2\pi)^2}\int d\phi \; [\,\cdot\,] \; ,
        \end{align}
        and $\phi$ is the polar angle in Fourier space. This is a data compression step from images to a useful representation each of size $(2,N_\text{bins})$, which acts as the data input to all subsequent part of the analysis. We can write
\begin{align}
	\mathbf{D}_{\bar\ell} &= \mathbf{D}^\text{rkSZ}_{\bar\ell} + \mathbf{D}^\text{tSZ}_{\bar\ell} + \cdots + \mathbf{D}^\text{noise}_{\bar\ell} \; ,
\end{align}
\begin{figure}
	\includegraphics[width=\linewidth]{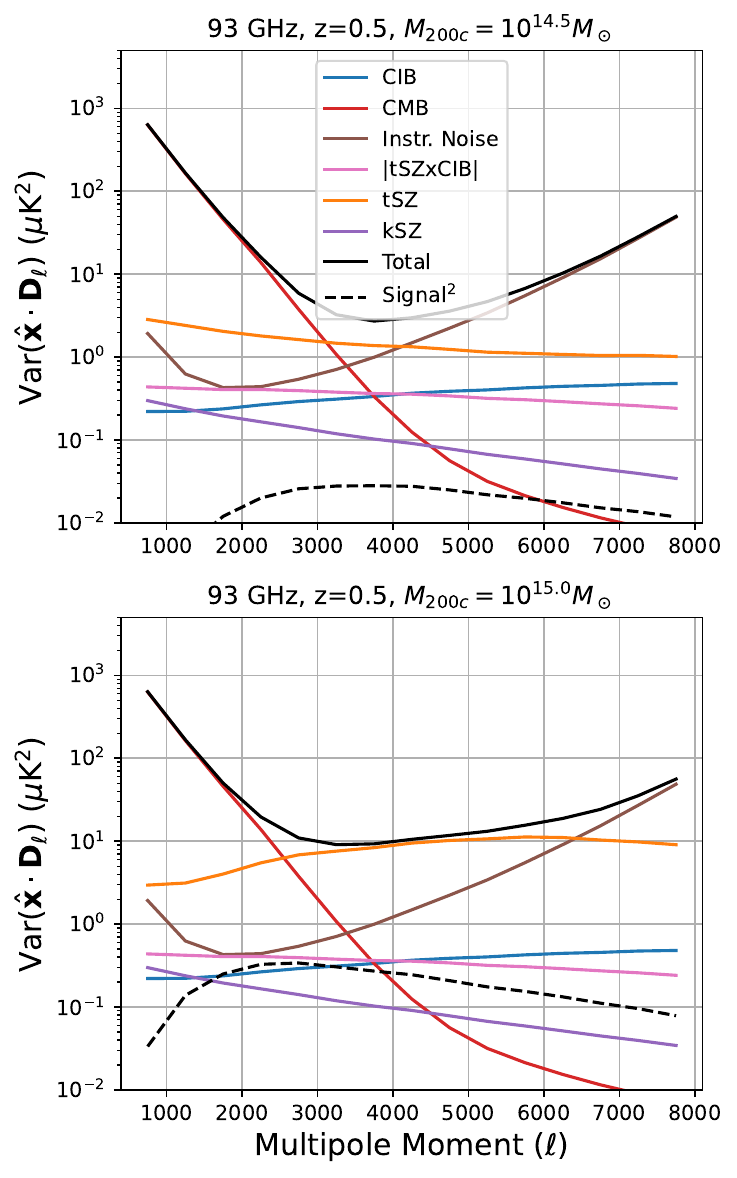}
	\caption{For the two cluster masses of $10^{14.5}$ (top) and $10^{15}\, M_\odot$ (bottom), the squared rkSZ signal $\mathcal{S}^2\bar{F}_{\bar\ell}^2$ (dashed black) is plotted along with the noise variance $\operatorname{Var}(\hat{\Bell}_x\cdot\mathbf{D}^\text{comp}_{\bar\ell})$ for the following components: CIB (blue), tSZ (orange), CMB (red), $|$tSZ$\times$CIB$|$ (pink), kSZ (purple) and beam-deconvolved SO instrumental noise (brown). The total noise variances are plotted as the solid black lines.
    For homogeneous and isotropic noise e.g.\ the instrumental noise, the noise variance agrees with analytical estimates from component power spectra in Eq.\ \eqref{eq:Cbin}. 
    }
	\label{fig:Dk_comps}
\end{figure}
where the ellipsis captures all other foregrounds one intends to model.
By restricting the estimator to $\operatorname{Im}[m(\Bell)]$, contaminants that are even under reflection about the cluster center e.g.\ the tSZ monopole, the bulk kSZ for a spherical gas distribution, are purely real in Fourier space and is nulled identically. Such components enter the analysis only through their miscentering and asphericity.

For the rkSZ signal:
\begin{align}
	\mathbf{D}^\text{rkSZ}_{\bar\ell} &= - |\hat{\mathbf{J}}\times\hat{\mathbf{n}}|  \mathcal{S} \int^{\bar\ell+\Delta \ell/2}_{\bar\ell - \Delta\ell/2} \frac{\ell d\ell}{(2\pi)^2} F(\ell) \int d\phi (\hat{\mathbf{d}}\cdot\hat{\Bell})  \hat{\Bell} \nonumber \\
					  &\equiv |\hat{\mathbf{J}}\times\hat{\mathbf{n}}| \mathcal{S} \bar{F}_{\bar\ell} \hat{\mathbf{d}} \; . \label{eq:for_appB_D}
\end{align}
We have absorbed the prefactor into the binned rkSZ Fourier profile, defined as
\begin{align}
	\bar{F}_{\bar\ell} &\equiv -\frac{1}{2} \int_{\bar\ell\text{-bin}} \frac{d^2\ell}{(2\pi)^2} F(\ell) \; .
\end{align}
The other (foreground and noise) components to $\mathbf{D}_{\bar\ell}$ can be estimated from simulations.

		\textbf{Step 2.}  Apply the optimal $\bar\ell$-weighting (matched filter) based on the data model to get a single 2D vector $\mathbf{V}$ per object:
		\begin{align}
			\mathbf{V} \equiv \sum_{\bar\ell} \Psi_{\bar\ell}\mathbf{D}_{\bar\ell} \; . \label{eq:for_appB_V}
		\end{align}
		The optimal filter is derived to provide the correct normalization $\langle\mathbf{V}  \rangle= \langle|\hat{\mathbf{J}}\times\hat{\mathbf{n}}|\rangle  \mathcal{S}\hat{\mathbf{d}}$ (assuming the noise has zero correlation with the dipole direction), and to minimize the variance $\operatorname{Var}(\hat{\mathbf{d}}\cdot\mathbf{V})$. See Appendix \ref{app:filter_vec} for a derivation. In the general case, we have to construct an azimuthal-weighted covariance (the direction $\hat{\Bell}_x$ is arbitrary, assuming that the noise and foreground contributions are isotropic),
\begin{align}
	\bar{R}_{\bar\ell \bar\ell'} &\equiv \operatorname{Cov}(\hat{\Bell}_x\cdot\mathbf{D}_{\bar\ell},\hat{\Bell}_x\cdot\mathbf{D}_{\bar\ell'}) \label{eq:for_appB_cov} \\
	&= \int_{\bar\ell\text{-bin}} \frac{d^2\ell }{(2\pi)^2}\int_{\bar\ell'\text{-bin}} \frac{d^2 \ell'}{(2\pi)^2}\nonumber \\
				     &\phantom{ASDASD}\times\cos\phi \operatorname{Cov}(\operatorname{Im}[m(\Bell)],\operatorname{Im}[m(\Bell')])\cos\phi' \; ,
\end{align}
which is a $N_\text{bins}\times N_\text{bins}$ matrix with inverse $\bar{R}^{-1}_{\bar\ell \bar\ell'}$. 

Figure~\ref{fig:Dk_comps} plots the components contributing to the diagonal of $\bar R_{\bar\ell\bar\ell'}$ at $z=0.5$. For observation at 93 GHz, the lensed CMB dominates at $\ell\lesssim2500$. Depending on mass, the cluster's own miscentered tSZ dominates the intermediate scales at $3000\lesssim\ell\lesssim4000$ (or 6000) for $M_{200c}=10^{14.5}$ ($10^{15}$)$\,M_\odot$, and the instrumental noise dominates at small scales. 
See Appendix \ref{app:cov} for the analytical modeling of the tSZ covariance.

In the case of isotropic homogeneous noise where  $\operatorname{Cov}(\operatorname{Im}[m(\Bell)],\operatorname{Im}[m(\Bell')])=(2\pi)^2C^\text{hom}_\ell[\delta^2(\Bell-\Bell')-\delta^2(\Bell+\Bell')]/2$, the covariance becomes diagonal:
\begin{align}
	\bar{R}_{\bar\ell\bar\ell'}
				       &= \pi  \delta_{\bar\ell\bar\ell'} \int^{\bar\ell+\Delta\ell/2}_{\bar\ell-\Delta\ell/2} \frac{\ell d\ell}{(2\pi)^2}  C_\ell^\text{hom} \; \\
				       &\equiv  \delta_{\bar\ell \bar\ell'} \bar{C}^\text{hom}_{\bar\ell} \; , \label{eq:for_appB_Chom}
\end{align}
where we absorbed the $1/2$ prefactor into the definition of the binned spectra,
\begin{align}
	\bar{C}^\text{hom}_{\bar\ell} \equiv \frac{1}{2} \int_{\bar\ell\text{-bin}} \frac{d^2\ell}{(2\pi)^2} C^\text{hom}_\ell \; .
	\label{eq:Cbin}
\end{align}

We provide two sets of filters: A.) optimal filter for general noise (where $\bar{R}_{\ell\ell'}$ is to be estimated from simulations), 
\begin{align}
	\Psi_{\bar\ell} &= N_\text{V}^2\sum_{\bar\ell'} \bar{F}_{\bar\ell'} \bar{R}^{-1}_{\bar\ell'\bar\ell} \\
	N_\text{V}^{-2} &= \sum_{\bar\ell,\bar\ell'}\bar{F}_{\bar\ell}\bar{R}^{-1}_{\bar\ell\bar\ell'} \bar{F}_{\bar\ell'} \; .
\end{align}
When the tSZ is not deprojected, one should note that variation in the covariance $\bar{R}_{\Bell\Bell'}$ will be significant as the tSZ power can exceed that of instrumental noise in the intermediate scales ($3000\lesssim\ell\lesssim6000$ for $M_{200c}=10^{15}\,M_\odot$ at $z=0.5$ from Fig.\ \ref{fig:Dk_comps}). Note that a mass- and redshift-dependent multifrequency matched filter is necessary for optimal detection (Fig.\ \ref{fig:R_mmf}; see next subsection). If one resorts to tSZ deprojection (which is close to optimal for the highest-mass clusters $M_{200c}\gtrsim 10^{15} M_\odot$, see Fig.\ \ref{fig:perobject_snr}) at the expense of a halved cumulative SNR (see Fig.\ \ref{fig:cum_snr}), one can apply the following filter on constrained ILC maps (with caution for clustered dust contribution on the covariance):

B.) Specializing the results above to homogeneous isotropic noise
		\begin{align}
			\Psi_{\bar\ell} &= N^2_\text{V} \bar{F}_{\bar\ell}/\bar{C}_{\bar\ell}^\text{hom}\label{eq:cilc_filter}  \\
	N_\text{V}^{-2} &\equiv  \sum_{\bar\ell} \bar{F}^2_{\bar\ell} /\bar{C}^\text{hom}_{\bar\ell} \label{eq:cilc_filter2} \; ,
		\end{align}
which is invalid in the presence of clustered noise e.g.\ tSZ, but can be applied together with a multifrequency deprojection of clustered components. Relevant expressions for the multifrequency matched filter will be presented in the next subsection.

\textbf{Step 3.}  The weighted data $\mathbf{V}(\hat{\mathbf{n}})$ as a function of the cluster centers $\hat{\mathbf{n}}$ is cross-correlated with external spin-1 fields. Cross-correlating $\mathbf{V}$ with the true dipole direction  $\hat{\mathbf{d}}$ with an optimal object-weighting maximizes the rkSZ signal:
\begin{align}
	\langle \hat{\mathbf{d}}\cdot\mathbf{V}  \rangle &= \langle|\hat{\mathbf{J}}\times\hat{\mathbf{n}}|\rangle \mathcal{S} \; , \\
\operatorname{Var}(\hat{\mathbf{d}}\cdot\mathbf{V}) &= N^2_\text{V} \; .
\end{align}

We do not consider the auto-statistic $\langle\mathbf{V}\cdot\mathbf{V}\rangle$: extracting $\mathcal{S}^2$ from it would require the total foreground-plus-noise variance $N^2_\text{V}$ to be known accurately and subtracted off, which is unrealistic for the clustered foregrounds considered here. All statistics below are therefore cross-correlations with external orientation information.

\subsection{Multifrequency Matched Filter expressions}
\begin{figure}
	\includegraphics[width=\linewidth]{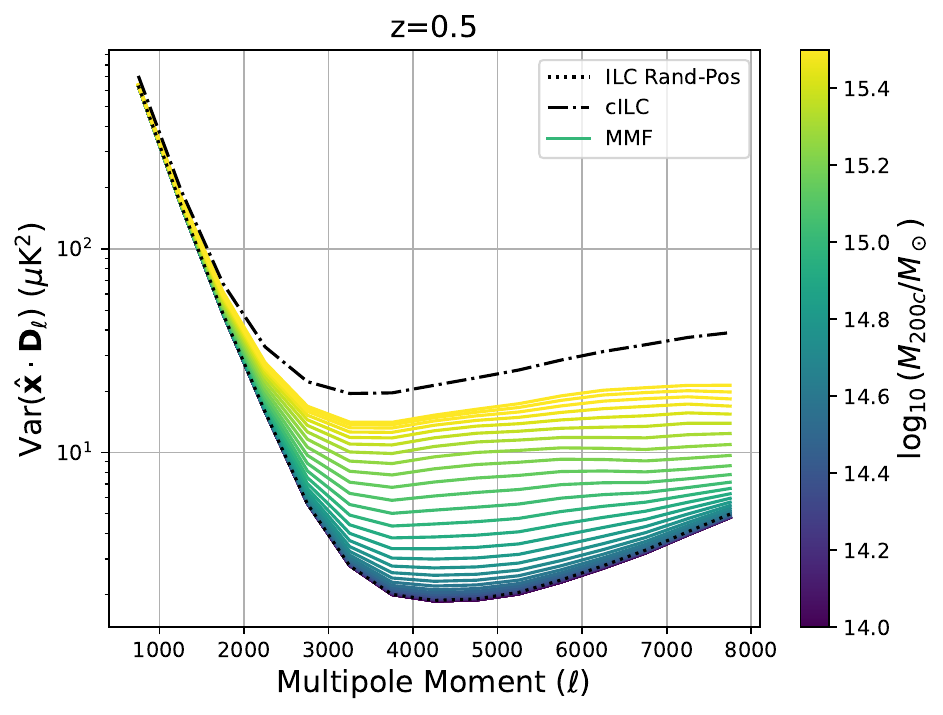}
	\caption{Effective filtered-data variance $\operatorname{Var}(\hat{\Bell}_x\cdot\mathbf{D}_{\bar\ell})$ at $z=0.5$ entering the MMF covariance, as a function of cluster mass (color scale), compared against the tSZ-deprojected cILC variance (dash-dotted) and the ILC variance evaluated at random positions (ILC Rand-Pos; dotted). At low mass the MMF variance approaches the ILC Rand-Pos level; with increasing mass the cluster's own tSZ raises the MMF-cleaned variance toward (but never above) the cILC level. This mass dependence of the covariance is what drives the mass dependence of the optimal filter.}
	\label{fig:R_mmf}
\end{figure}
The starting point of the preceding subsection are single-frequency CMB maps $m(\Bell)$---these single-frequency maps could be ILC or constrained ILC cleaned maps. This subsection combines both multifrequency cleaning and matched filtering into one single multifrequency matched filter (MMF) $\Psi^\mu_\ell$, where $\mu$ denotes the frequency channel. The key motivation for this generalization is that tSZ power varies drastically with mass across cutouts centered at clusters, and is the dominant contribution of power  for the highest-mass clusters. Using ILC weights---which derive from full-sky multifrequency power spectra---actually yields higher variance than using cILC weights due to the significant suboptimality. 
Therefore the MMF $\Psi^\mu_\ell$ is mass and redshift dependent. 

Figure~\ref{fig:R_mmf} demonstrates the mass-dependence of the MMF-filtered noise: at $\ell\lesssim2000$ the filtered-data
variance is set by the isotropic primary CMB and is the same for all masses, whereas at
smaller scales the cluster's own tSZ varies with mass, changing the optimal weights and variance compared with that obtained for random sky positions (ILC rand-pos), while staying below the
tSZ-deprojected cILC variance. 
We repeat the steps of the last subsection to construct the multifrequency matched filters:

\textbf{Step 1.}
The data vectors are promoted to multifrequency data vectors:
\begin{align}
	\mathbf{D}^\mu_{\bar\ell} \equiv
						 \int_{\bar{\ell}\text{-bin}} \frac{d^2\ell}{(2\pi)^2} \operatorname{Im}[m^\mu(\Bell)]\hat{\Bell} \; . 
\end{align}
In terms of the blackbody SED $a^\mu$, tSZ SED  $b^\mu$, etc., 
 \begin{align}
	 \mathbf{D}^\mu_{\bar\ell} = a^\mu \mathbf{D}^\text{rkSZ}_{\bar\ell} + b^\mu \mathbf{D}^\text{tSZ}_{\bar\ell} + \cdots + \mathbf{D}^{\mu,\text{noise}}_{\bar\ell} \; .
\end{align}
\textbf{Step 2.}
Optimizing also over frequency, we get the MMF:
\begin{align}
	\mathbf{V} \equiv \Psi^\mu_{\bar\ell} \mathbf{D}_{\bar\ell}^\mu \equiv \Psi_{\bar{L}} \mathbf{D}_{\bar{L}} \; , \label{eq:for_appB_multiV}
\end{align}
where $\bar{L}=(\mu,\bar\ell)$ is a compact index of both frequency and scale. The estimator should normalize to $\langle a^\mu \Psi^\mu_{\bar\ell} \mathbf{D}^\text{rkSZ}_{\bar\ell}\cdot\hat{\mathbf{d}}  \rangle = \langle|\hat{\mathbf{J}}\times\hat{\mathbf{n}}|\rangle \mathcal{S}$. 
This corresponds to the constraint
\begin{align}
	\Psi^\mu_{\bar\ell} a^\mu  \bar{F}_{\bar\ell} \equiv \Psi_{\bar{L}} \bar{F}_{\bar{L}}= 1 \; \quad \bar{F}_{\bar{L}}\equiv a^\mu \bar{F}_{\bar\ell} \; . \label{eq:for_appB_multiD}
\end{align}
The multifrequency data covariance is
\begin{align}
	\bar{R}^{\mu\nu}_{\bar\ell \bar\ell'} &\equiv \operatorname{Cov}(\hat{\Bell}_x\cdot\mathbf{D}^\mu_{\bar\ell},\hat{\Bell}_x\cdot\mathbf{D}^\nu_{\bar\ell'})\label{eq:for_appB_multicov} \\
	&= \int_{\bar\ell\text{-bin}} \frac{d^2\ell }{(2\pi)^2}\int_{\bar\ell'\text{-bin}} \frac{d^2 \ell'}{(2\pi)^2}\nonumber \\
				     &\phantom{ASDASD}\times\cos\phi \operatorname{Cov}(\operatorname{Im}[m^\mu(\Bell)],\operatorname{Im}[m^\nu(\Bell')])\cos\phi' \; ,
\end{align}
The MMF is obtained straightforwardly (App. \ref{app:filter_vec}) as
\begin{align}
	\Psi_{\bar{L}} &= N_\text{V}^2 \bar{F}_{\bar{L}'} \bar{R}^{-1}_{\bar{L}'\bar{L}} \; , \\
	N_\text{V}^{-2} &\equiv \bar{F}_{\bar{L}} \bar{R}^{-1}_{\bar{L}\bar{L}'} \bar{F}_{\bar{L}'} \; .
\end{align}
\textbf{Step 3.} It follows again
\begin{align}
	\langle \hat{\mathbf{d}}\cdot\mathbf{V}  \rangle &= \langle|\hat{\mathbf{J}}\times\hat{\mathbf{n}}|\rangle \mathcal{S}\; , \\
	\operatorname{Var}(\hat{\mathbf{d}}\cdot\mathbf{V}) &= N^2_\text{V} \; .
\end{align}
The challenge to this method lies in determining the tSZ contribution in the mass-redshift dependent $\bar{R}_{\bar{L}\bar{L}'}$ (see App.\ \ref{app:cov}). Realistically it should capture the variation in profile, asphericity, and miscentering of the tSZ component. Our modeling of $\bar{R}$ considers the miscentering of a spherical tSZ profile at a mass and redshift, thus ignores the internal scatter of the tSZ. While a mismodeled covariance do not bias the estimator, a Monte-Carlo estimation from the filtered vectors $\mathbf{V}$ would remove the modeling uncertainty from entering the signal-to-noise estimation.

\subsection{Recovering the rkSZ dipole (headless) axis: the tensor matched filter formalism}
The dipole signal can be manipulated into a quadrupole by squaring the imaginary part of the Fourier map and subtracting the square of a $90^\circ$-rotated copy, where the rotated map $\operatorname{Im}^T[m(\Bell)]$ is defined as
 \begin{align}
	 \operatorname{Im}^T[m(\Bell)] &\equiv \operatorname{Im}[m(\Bell^T)] \\
	 \Bell^T &\equiv \Bell\times\hat{\mathbf{n}} =
	 \begin{pmatrix}
	 	\ell_y \\ -\ell_x
	 \end{pmatrix} \; .
\end{align}
 Upon this manipulation, the rkSZ signal takes a quadrupolar form:
\begin{align}
	\operatorname{Im}[\bar{\Theta}_\text{rkSZ}&(\Bell)]^2 - \operatorname{Im}^T[\bar{\Theta}_\text{rkSZ}(\Bell)]^2 \nonumber \\
	&= |\hat{\mathbf{J}}\times\hat{\mathbf{n}}|^2  \mathcal{S}^2\left[(\hat{\mathbf{d}}\cdot\hat{\Bell})^2-(\hat{\mathbf{d}}\cdot\hat{\Bell}^T)^2 \right]F^2(\ell)  \nonumber \\
	&= |\hat{\mathbf{J}}\times\hat{\mathbf{n}}|^2  \mathcal{S}^2\cos(2(\phi-\phi_0))F^2(\ell) \; ,
\end{align}
where $\phi_0$ is the polar angle of the true dipole direction.
Note that for statistically isotropic noise $\langle\operatorname{Im}[n]^2\rangle=\langle\operatorname{Im}^T[n]^2\rangle$, so the rotation-subtraction also removes the noise bias of the quadratic statistic.
The data analysis methodology can be interpreted in three steps:

\textbf{Step 1.} Define the tensor
\begin{align}
	\mathbb{D}^{jk}_{\bar\ell} &\equiv \int_{\bar\ell\text{-bin}} \frac{d^2\ell}{(2\pi)^2}  \left[\operatorname{Im}[m(\Bell)]^2 - \operatorname{Im}^T[m(\Bell)]^2 \right] \hat{\ell}_j\hat{\ell}_k \; ,
\end{align}
and the dipole axis tensor
\begin{align}
	e_{jk} &\equiv \hat{\mathbf{d}}_j\hat{\mathbf{d}}_k - (\hat{\mathbf{n}}\times\hat{\mathbf{d}})_j(\hat{\mathbf{n}}\times\hat{\mathbf{d}})_k \label{eq:dipole_tensor}\\
	&=
	\begin{pmatrix}
		\hat{d}_x^2-\hat{d}_y^2 & 2\hat{d}_x\hat{d}_y \\
		2\hat{d}_x \hat{d}_y & -\hat{d}_x^2 + \hat{d}_y^2
	\end{pmatrix} = 
	\begin{pmatrix}
		\cos 2\phi_0 & \sin 2\phi_0 \\
		\sin 2\phi_0 & -\cos 2\phi_0
	\end{pmatrix} \; .
\end{align}
For the rkSZ$\times$rkSZ signal part, where we substitute $m(\Bell)=-i  |\hat{\mathbf{J}}\times\hat{\mathbf{n}}|  \mathcal{S}\hat{\mathbf{d}}\cdot\hat{\Bell} F(\ell)$,
\begin{align}
	\mathbb{D}^{jk,\text{rkSZ}}_{\bar\ell} &=  |\hat{\mathbf{J}}\times\hat{\mathbf{n}}|^2 \mathcal{S}^2 \int_{\bar\ell\text{-bin}} \frac{d^2\ell}{(2\pi)^2} F^2(\ell) \cos(2(\phi-\phi_0))\hat{\ell}_j \hat{\ell}_k \nonumber \\
					       &\equiv  |\hat{\mathbf{J}}\times\hat{\mathbf{n}}|^2  \frac{1}{2} \mathcal{S}^2 e_{jk} \bar{G}_{\bar\ell} \; ,
\end{align}
where
\begin{align}
	\bar{G}_{\bar\ell} &\equiv \frac{1}{2}  \int_{\bar\ell\text{-bin}} \frac{d^2\ell}{(2\pi)^2} F^2_\ell  \; ,
\end{align}
therefore cross-correlating the data tensor $\mathbb{D}$ with the dipole axis tensor  $e_{jk}$ recovers the rkSZ signal.

\textbf{Step 2.} Apply the matched filtering to get a tensor $\mathbb{Q}$ for each object:
 \begin{align}
	 \mathbb{Q} \equiv \Psi_{\bar\ell}\mathbb{D}_{\bar\ell}^{jk} \; .
\end{align}
The data covariance (evaluating along an arbitrary axis, as we expect the noise and foreground to be statistically isotropic, here $e'_{jk}$ taken as the headless $x$-axis), is
\begin{align}
	\bar{R}_{\bar\ell\bar\ell'} &= \operatorname{Cov}(e_{jk}\mathbb{D}^{jk}_{\bar\ell},e_{jk}\mathbb{D}^{jk}_{\bar\ell'})  \\
	&= \int_{\bar\ell\text{-bin}} \frac{d^2\ell }{(2\pi)^2}\int_{\bar\ell'\text{-bin}} \frac{d^2 \ell'}{(2\pi)^2} \times \cos2\phi \cos2\phi' \nonumber \\
	&\phantom{ASDADSASAS}\times\operatorname{Cov}( \operatorname{Im}[m(\Bell)]^2 - \operatorname{Im}^T[m(\Bell)]^2, \nonumber \\
	&\phantom{ASDASDAAS'\times \operatorname{Cov}( }
	\operatorname{Im}[m(\Bell')]^2-\operatorname{Im}^T[m(\Bell')]^2) \\
	&=4 \int_{\bar\ell\text{-bin}} \frac{d^2\ell }{(2\pi)^2}\int_{\bar\ell'\text{-bin}} \frac{d^2 \ell'}{(2\pi)^2} \times \cos2\phi \cos2\phi' \nonumber \\
	&\phantom{ASDADSASAS}\times\operatorname{Cov}( \operatorname{Im}[m(\Bell)]^2,\operatorname{Im}[m(\Bell')]^2) \; ,
\end{align}
the filter has the same functional form of
\begin{align}
	\Psi_{\bar\ell}	 &= N_\text{T}^2  \bar{R}^{-1}_{\bar\ell\bar\ell'}\bar{G}_{\bar\ell'} \\
	N_\text{T}^{-2} &= \bar{G}_{\bar\ell}\bar{R}^{-1}_{\bar\ell\bar\ell'}\bar{G}_{\bar\ell'} \;.
\end{align}

\textbf{Step 3.} The tensor $\mathbb{Q}$ is to be cross-correlated with an external tensor field, such as the shape field of clusters, or the large-scale tidal field. Cross-correlating with the true dipole axis gives
 \begin{align}
	 \langle e_{jk}\mathbb{Q}^{jk}  \rangle &=  \langle|\hat{\mathbf{J}}\times\hat{\mathbf{n}}|^2\rangle\mathcal{S}^2 \; , \\
	 \operatorname{Var}(e_{jk}\mathbb{Q}^{jk}) &= N^2_\text{T} \; .
\end{align}
\section{SNR Forecast}
\label{sec:forecast}
\begin{figure*}
	\includegraphics[width=0.95\linewidth]{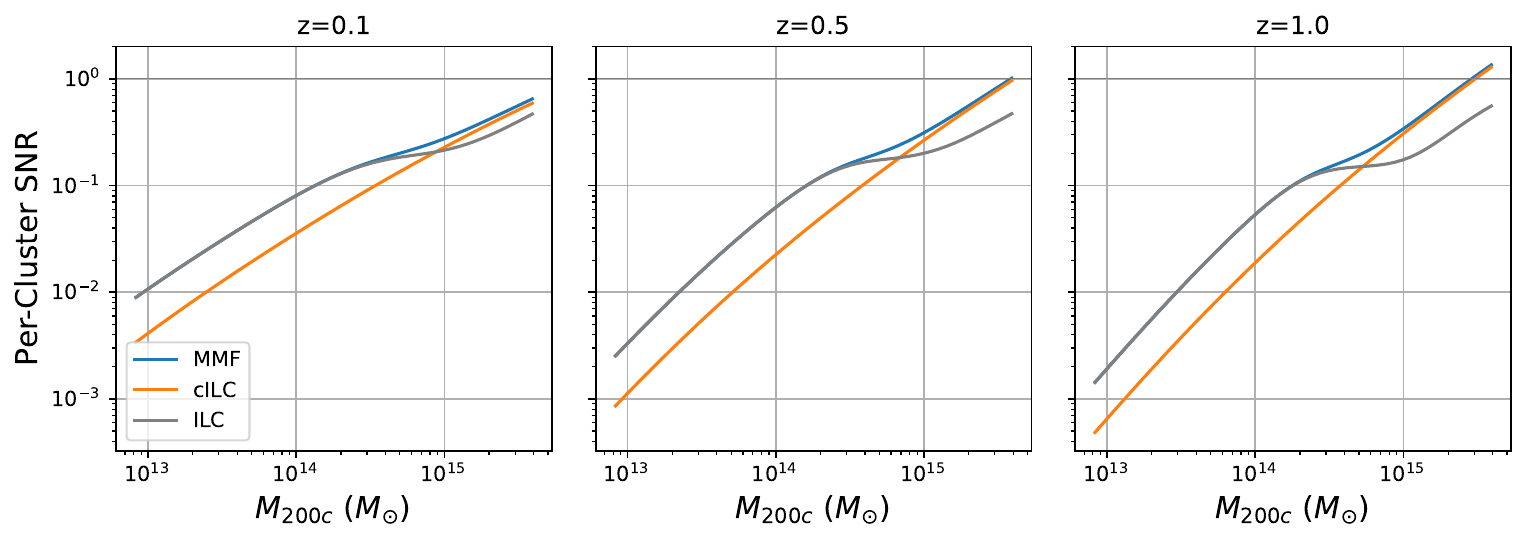}
	\caption{The estimated SNR for the rkSZ effect for a cluster of mass $M_{200c}$ and redshift  $z=0.1$ (left),  $z=0.5$ (middle), and  $z=1.0$ (right) for SO using the vector cross-correlation estimator (with $f_\text{misalign}=1$). For the optimal MMF (blue) and with (orange) tSZ deprojection, only the highest-mass clusters of mass $\gtrsim 10^{15} M_\odot$ could reach a single-object SNR of above 1, while the ILC filter (gray), whose weights derive from full-sky power spectra, underestimates the cluster-centered tSZ and is suboptimal at the high-mass end. At masses of $M_{200c}\lesssim 10^{14} M_\odot$, the MMF converges to the ILC solution. We take a cosmological average value for the spin parameter, $\lambda_\text{gas}=$0.046. }
	\label{fig:perobject_snr}
	\subfloat[MMF\label{fig:perobject_snr_a}]{
    \includegraphics[trim={0.0cm 0.0cm 1cm 1.2cm}, clip , width=0.48\textwidth]{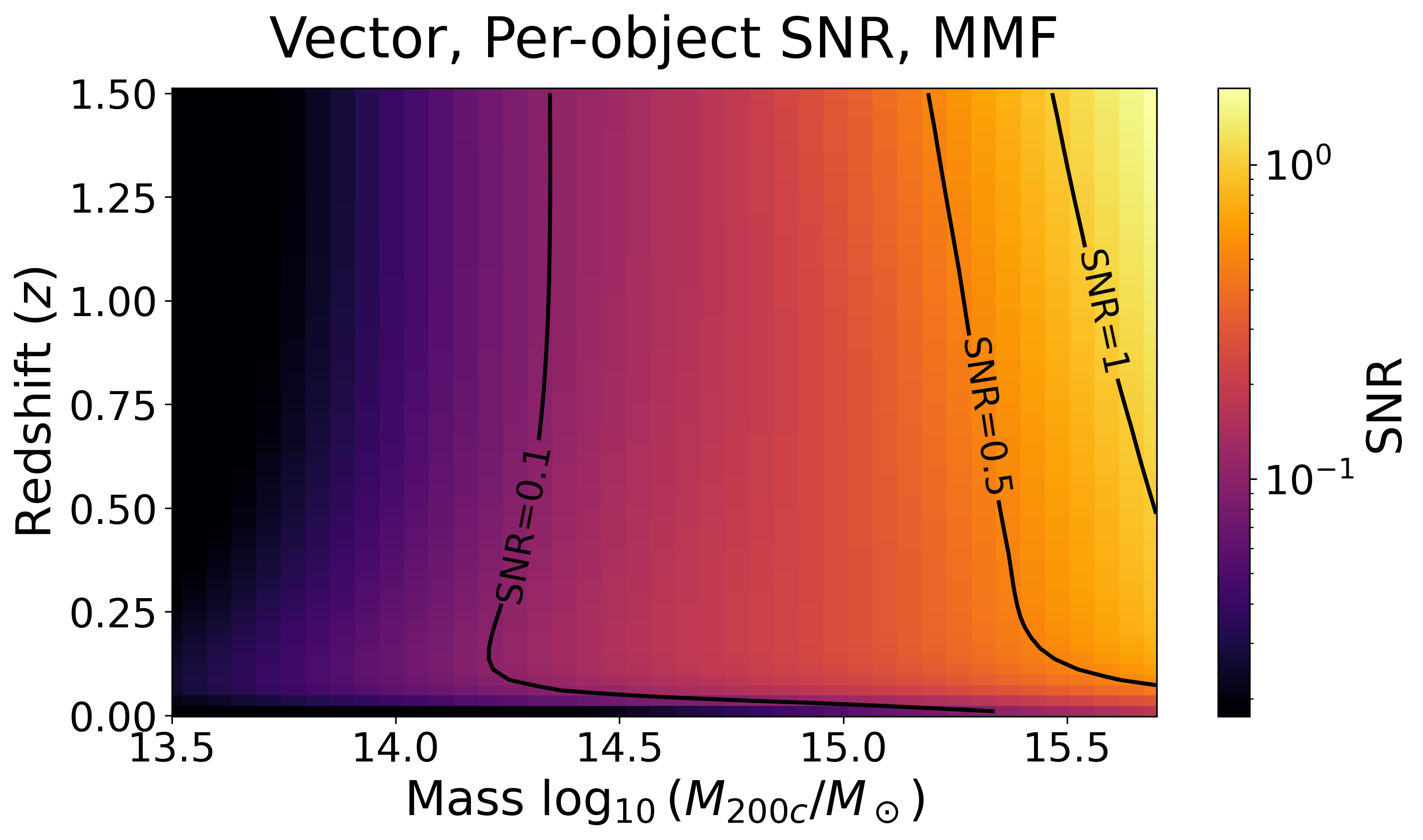}
    }
    \hfill
    \subfloat[tSZ-deprojected cILC\label{fig:perobject_snr_b}]{
        \includegraphics[trim={1cm 0.0cm 0.0cm 1.2cm}, clip, width=0.48\textwidth]{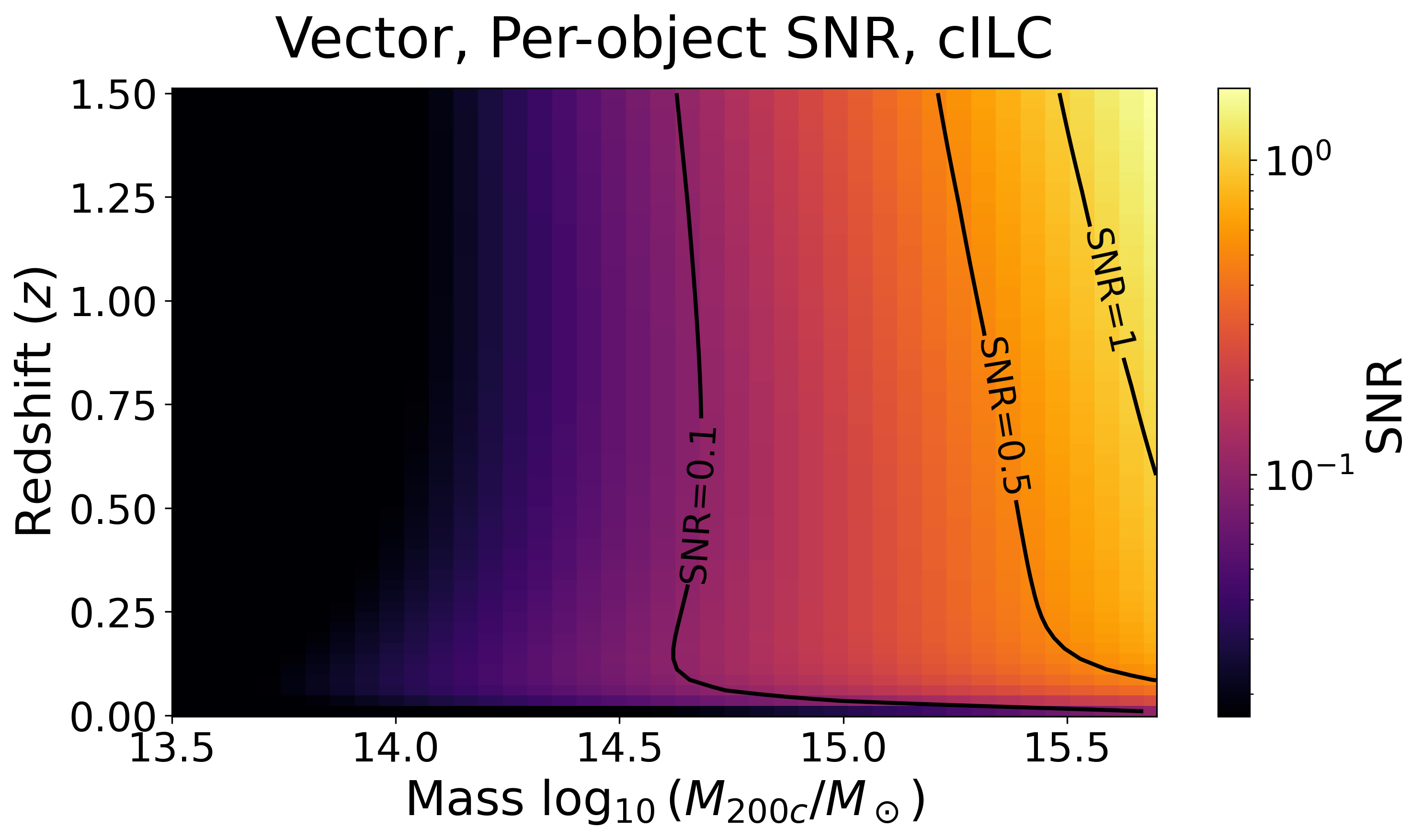}
    }
	\caption{Similar to Fig.\ \ref{fig:perobject_snr}, but we plot the single-object SNR of the vector cross-correlation estimator as a function of mass and redshift for the optimal MMF (left) and with the tSZ deprojected cILC (right).}
	\label{fig:perobject_snr2}
\end{figure*}
We apply the matched-filter formalism above on mock observations. The mock data is constructed from simulated component sky maps, forward-modeled to data cutouts in real-space then Fourier-space, to the (multifrequency) data vectors $\mathbf{D}_{\bar\ell}$.
	We use simulated data up to $\ell_\text{max}=8000$ with bin size of $\Delta \ell = 500$.

Our forecast has the following specifications:
\begin{itemize}
	\item Survey specification: for our baseline we generate Gaussian noise with SO (goal) noise spectra \cite{so19_whitepaper}. A systematic exploration of other experimental configurations (ACT, \textit{Planck}, CMB-HD, and their combinations) is deferred to future work.

	\item Filter: We construct our matched filter in two ways. 1.\ The \texttt{MMF} mode correspond to a mass-redshift dependent multifrequency matched filter where the covariance $\bar{R}_{\bar\ell\bar\ell'}$ is modeled from simulations and gNFW fits. 2.\ the \texttt{cILC} mode is an implementation of the simple mass-redshift independent filter in Eq.\ \ref{eq:cilc_filter} which inputs the tSZ-deprojected full-sky power spectra in the covariance $\bar{R}_{\bar\ell\bar\ell'}$.
	\item Analytical vs Monte-Carlo cumulative SNR estimates: from the objects $\mathcal{S}$ and  $N^2_\text{V/T}$ from the last section, we can compute an analytical estimate of the SNR. This assumes that the covariance is modeled correctly. While systematic errors on the covariance do not bias the matched filtering, the forecasted SNR can be overestimated. We therefore develop a simple Monte-Carlo scheme in a way similar to actual observations (e.g.\ Ref.\ \cite{goldstein25_rksz}) to determine the cumulative SNR of all clusters combined.
    \item Mass range: To minimize the baryonic-feedback uncertainty, we only utilize the mass range of $\gtrsim 10^{14}\, M_\odot$ for our SNR forecast.
    Although in Sec.\ \ref{sec:rksz} we compared our model with TNG-Cluster and TNG300-1 down to $10^{13.5}\,M_\odot$ at $z=0$, we consider forecast results using clusters of mass $M_{200c}\lesssim 10^{14}\, M_\odot$ as extrapolations. See Sec.~\ref{sec:discussion} for a discussion of baryonic feedback.
\end{itemize}

\subsection{Analytical Single-Object SNR}
For a single object (labeled $i$), the per-object SNR estimate for the detection of the cross-correlation of the rkSZ signal with an external proxy $\hat{\mathbf{d}}^\text{proxy}$ (that is misaligned with the true dipole direction by $f_\text{misalign}\equiv \hat{\mathbf{d}}\cdot\hat{\mathbf{d}}^\text{proxy}$, the cosine of the misalignment angle) is 
\begin{align}
	\mathrm{SNR}_{\text{V},i} &\equiv \frac{ f_{\text{misalign},i}  \langle |\hat{\mathbf{J}}\times\hat{\mathbf{n}}|  \rangle   \mathcal{S}_i}{{N_{\text{V},i}}}  \\
	\mathrm{SNR}_{\text{T},i} 
				     &\equiv \frac{ (2f^2_{\text{misalign},i}-1)  \langle |\hat{\mathbf{J}}\times\hat{\mathbf{n}}|^2  \rangle \mathcal{S}^2_i}{{N_{\text{T},i}}}  \; .
\end{align}
Combining objects with the following weights
\begin{align}
	W^\text{V}_i &= \frac{f_{\text{misalign},i}  \langle |\hat{\mathbf{J}}\times\hat{\mathbf{n}}|  \rangle \mathcal{S}_i}{N^2_{\text{V},i}}\; , \\ 
	W^\text{T}_i &= \frac{ (2f^2_{\text{misalign},i}-1) \langle |\hat{\mathbf{J}}\times\hat{\mathbf{n}}|^2  \rangle  \mathcal{S}^2_i}{N^2_{\text{T},i}}\; ,
	\label{eq:combine_weights}
\end{align}
maximizes the catalog-combined SNR:
\begin{align}
	\text{SNR}^2_{\text{V/T,tot}} 
	 &= \sum_i \text{SNR}^2_{\text{V/T},i} \; .
	 \label{eq:cum_snr}
\end{align}
If the external proxy contains information on $f_{\text{misalign},i}$ and the rkSZ amplitude  $\mathcal{S}_i$, they should be substituted into Eq.\ \eqref{eq:cum_snr}. 
We present forecast for $f_{\text{misalign},i}=1$, corresponding to known spin direction for every object. 

In Figs.\ \ref{fig:perobject_snr} and \ref{fig:perobject_snr2}, we show the single-object SNR as a function of mass and redshift. At the highest-mass end ($M_{200c}\gtrsim 10^{15}M_\odot$), the clustered tSZ dominates the estimator variance and the optimal \texttt{MMF} converges to the \texttt{cILC} mode as the optimal solution converge to identically removing the tSZ component through deprojection; meanwhile the \texttt{ILC} filter constructed from full-sky power substantially underestimates the clustered tSZ and becomes suboptimal. At $M_{200c}\lesssim 10^{14.5} M_\odot$, the clustered tSZ power falls below the instrumental noise at small scales (see Fig.\ \ref{fig:Dk_comps}) and becomes sub-dominant, such that the \texttt{MMF} converge to the \texttt{ILC} solution as the dominant noise contribution (CMB, instrumental noise) are isotropic, whereas the \texttt{cILC} mode becomes suboptimal as noise penalty is paid to deproject the subdominant tSZ contribution. 
At higher redshift the \texttt{MMF} converges to the \texttt{cILC} mode at
lower mass (Fig.~\ref{fig:perobject_snr}), as they are more compact on the sky, and thus the tSZ dominates over the small-scale isotropic noise at a lower mass.

For a realistic external catalog of rotation direction or axis proxy with a mass-redshift distribution $P_\text{proxy}(M,z)$, the catalog-combined SNR is 
\begin{align}
	\text{cSNR}^2_{\text{V/T}} = N_\text{clus}\int dM dz P_\text{proxy}(M,z) \text{SNR}^2_\text{V/T}(M,z) \; ,
\end{align}
where $N_\text{clus}$ is the number of clusters in the external catalog.
The catalog-combined SNR also depends on whether it selects fast-rotating clusters, in which case the amplitude $\mathcal{S}$ modeled from the median will have to be replaced by that of the selected subpopulation ($\mathcal{S}\propto\lambda_\text{gas}$ of the selection.)
We study below a specific example of a mass-complete catalog.
\subsection{Cumulative SNR for Mass-Complete Catalog}
\label{sec:info_den}
\begin{figure*}
   \centering
    \subfloat[Vector\label{fig:cum_snr_a}]{
    \includegraphics[trim={0.0cm 0.0cm 0.0cm 0.8cm}, clip , width=0.48\textwidth]{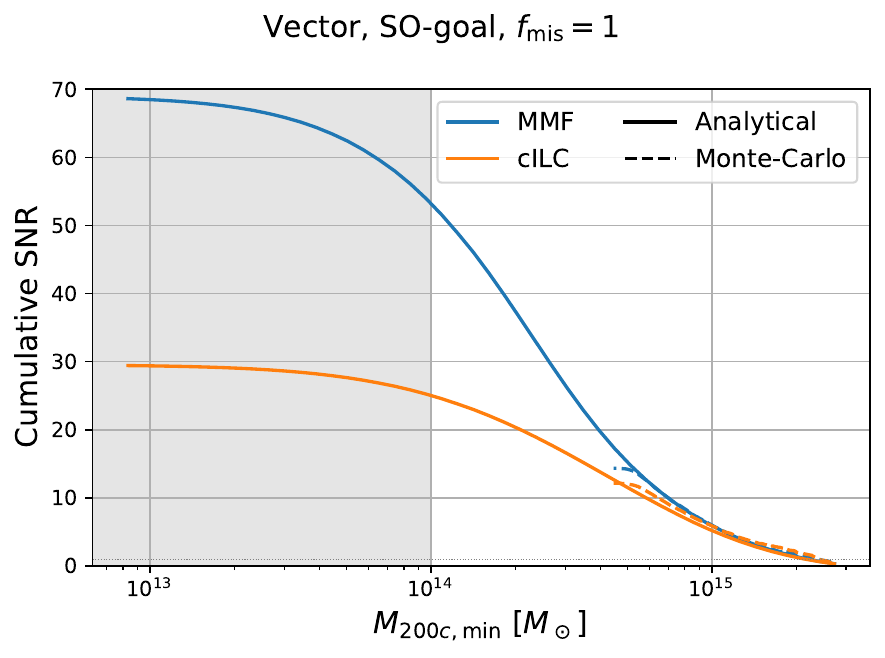}
    }
    \hfill
    \subfloat[Tensor\label{fig:cum_snr_b}]{
        \includegraphics[trim={0.0cm 0.0cm 0.0cm 0.8cm}, clip, width=0.48\textwidth]{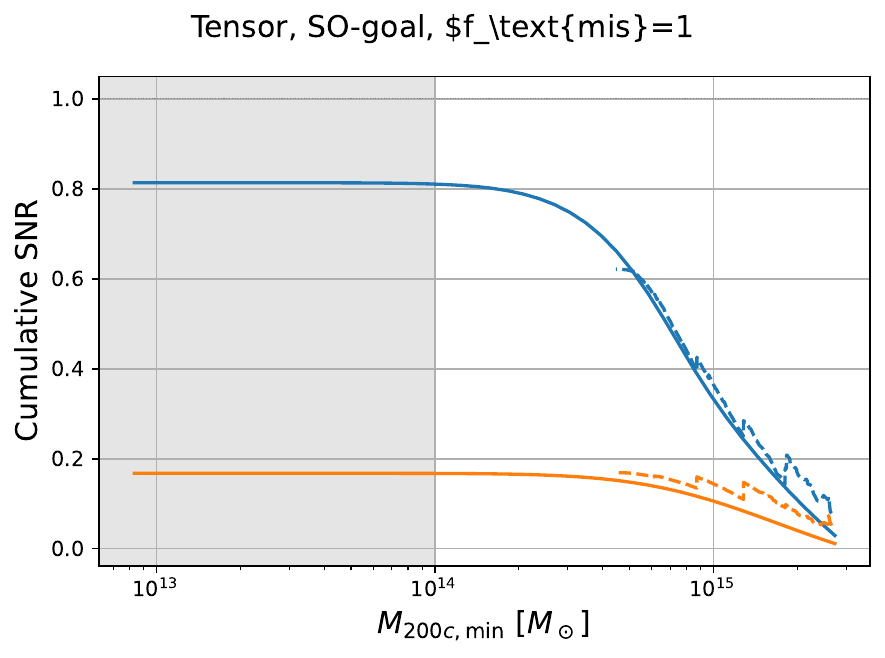}
    }
	\caption{The cumulative cross-correlation SNR assuming a mass-limited cluster sample (down to $M_\text{min}$, $x$-axis) with $f_\text{sky}=0.4$ and SO specifications using a vector (left) and a tensor (right) cross-correlation estimator, filtered using the optimal MMF (blue) and tSZ-deprojected cILC (orange). The weighting for each object assumes the amplitude of each rkSZ dipole to be $\mathcal{S}(M,z)$---the mean model amplitude at that mass and redshift, as well as perfect knowledge of the dipole direction $\hat{\mathbf{d}}$ from an external proxy, $f_\text{misalign}=1$. Multiply with $f_\text{misalign}$ estimated from the external proxy to obtain a realistic SNR forecast.
    Our forecast includes only clusters in the validated range of $M_{200c}\geq10^{14}\,M_\odot$ (non-shaded region), to limit the sensitivity to baryonic feedback in group-scale haloes (see Sec.~\ref{sec:discuss_sim_diff}).}
	\label{fig:cum_snr}
	\subfloat[MMF\label{fig:info_intensity_a}]{
    \includegraphics[trim={0.0cm 0.0cm 15cm 2cm}, clip , width=0.48\textwidth]{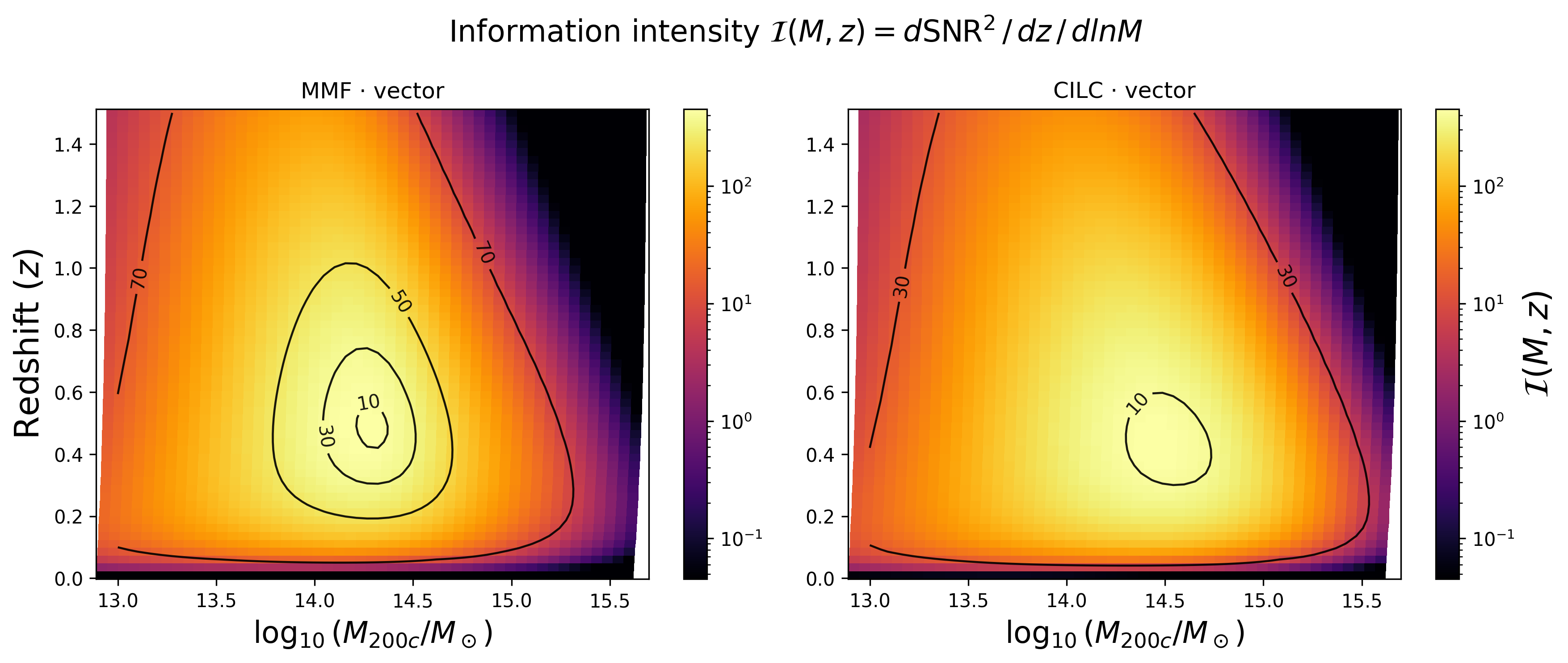}
    }
    \hfill
    \subfloat[tSZ-deprojected cILC\label{fig:info_intensity_b}]{
        \includegraphics[trim={15cm 0.0cm 0.0cm 2cm}, clip, width=0.48\textwidth]{./figs/info_intensity.png}
    }
	\caption{The information density for the vector cross-correlation estimator, $\mathcal{I}_\text{V}(M,z) = d^2\text{SNR}^2/dz/d\ln M$, described in Sec.\ \ref{sec:info_den} (Eq.\ \eqref{eq:info_density}, with  $f_\text{misalign}=1$), is plotted in the $(M,z)$ plane for the MMF (left) and the tSZ-deprojected cILC (right). The total SNR$^2$ contributed by a $(M,z)$-region is obtained by integrating the information density over that region. Two competing effects determine the information density: 1.\ the single-object SNR, which is a strong function of mass, and 2.\ the halo-mass function, which falls exponentially above $M_{200c}\sim 10^{14} M_\odot$.}
	\label{fig:info_den}
\end{figure*}
We present the cumulative SNR for a \emph{mass-complete} catalog (containing all clusters within a footprint of sky fraction $f_\text{sky}$ above  $M_\text{min}$).

Promoting the discrete catalog sum to a continuous integral over the halo mass function,
\begin{align}\label{eq:snr_integral} 
	\mathrm{cSNR}_{\text{V/T}}^2 &= 4\pi f_{\mathrm{sky}} \int dz \frac{d\chi}{dz} \chi^2 \int dM \frac{dn}{dM}  \mathrm{SNR}_{\text{V/T}}^2(M,z) \; ,
\end{align}
where $dn/dM$ is the halo mass function (HMF),
$\chi(z)$ is the comoving distance,
$\chi^2 \,d\chi/dz$ is the comoving volume per redshift per solid angle. We adopt the Tinker HMF with Planck~2018 cosmology \cite{tinker08_hmf}, evaluated using \texttt{colossus} \cite{diemer18_hmf}.

The halo model forecast integral (Eq.\ \eqref{eq:snr_integral}) sums contributions from all masses and redshifts. It is useful to identify which regions of the $(M,z)$ plane dominate the total SNR, both to guide observational strategy and to assess the robustness of the forecast to modeling assumptions. We define the \emph{information density}
\begin{align}
	\mathcal{I}_\text{V/T}(M,z) &\equiv \frac{d^2(\mathrm{cSNR}_\text{V/T}^2)}{d\ln M\,dz}\\
				    &=4\pi f_{\mathrm{sky}} \chi^2\frac{d\chi}{dz} M\frac{dn}{dM} \mathrm{SNR}_\text{V/T}^2(M,z) \; , \label{eq:info_density}
\end{align}
and the redshift-marginalized information density $\mathcal{I}(M)$,
 \begin{align}
	 \mathcal{I}_\text{V/T}(M) \equiv \int dz\, \mathcal{I}_\text{V/T}(M,z) \; ,
\end{align}
so that if we integrate the cumulative SNR down to a minimum halo mass $M_{200c}$, denoted  $M_\text{min}$, we have
\begin{align}
	\mathrm{cSNR}_\text{V/T}^2(M_\text{min})
						 &=  \int_{M_\text{min}}^{M_\text{max}} d\ln M\int dz\,\mathcal{I}_\text{V/T}(M,z) \; ,
\end{align}
where $M_\text{max}$ is arbitrary. The quantity $\mathcal{I}$ is the rkSZ analog of the $I_{\mathrm{patch}}$ diagnostic introduced in Ref.\ \cite{hotinli21_filter}. 
Expanding the single-object contribution,
\begin{align}
	\mathcal{I}_{\text{V}}(M,z) =4\pi {f_{\mathrm{sky}}} \chi^2\frac{d\chi}{dz} M\frac{dn}{dM} \frac{f^2_\text{misalign}  \langle |\hat{\mathbf{J}}\times\hat{\mathbf{n}}|  \rangle^2 \mathcal{S}^2(M,z)}{N^2_\text{V}(M,z)} \; .
\end{align}
Two competing effects shape $\mathcal{I}$:
\begin{itemize}
	\item \textbf{Signal growth}: the signal amplitude increases with mass as approximately $M_{200c}^{4/3}$ (Eq.\ \eqref{eq:sda_scaling}, and Fig.\ \ref{fig:montero_dorta}), and decrease with redshift as $D^{-2}_A$.
	\item \textbf{Abundance decline}: 
		$M\,dn/dM$ falls as a power law which steepens into an exponential decline at $\sim 10^{14}M_\odot$. This transition shifts to lower mass with increasing redshift, which sets the high-mass edge of the contours in Fig.\ \ref{fig:info_den}.
\end{itemize}
Their product determines a \emph{peak mass} $M_{\mathrm{peak}}(z)$ at which $\mathcal{I}$ is maximized for a given redshift. The peak mass is set by the competition between signal strength and halo rarity.
Figure \ref{fig:cum_snr} shows the resulting cumulative SNR as a function of the mass limit, and Fig.\ \ref{fig:info_den} the information density that builds it up. Three features anticipate the strategy discussion of Sec.\ \ref{sec:strategies}. (i) The MMF reaches $\mathrm{cSNR}\approx 53, 25$ for $M_\mathrm{min}=10^{14}, 10^{14.5}\,M_\odot$ respectively, of which only $\approx6$ comes from $M_{200c}>10^{15}\,M_\odot$: most of the information lives in the abundant, instrument-noise-dominated population, not in the individually loudest objects.  (ii) Applying tSZ deprojection uniformly across the catalog (cILC) forfeits roughly half of the total, $\mathrm{cSNR}\approx25$ at $M_\text{min}=10^{14}\, M_\odot$, because it pays an unnecessary noise penalty exactly where the information peaks. (iii) The information density peaks at $M_{200c}\approx10^{14.2}\,M_\odot$ and $z\approx0.4$--$0.6$, set by the signal-growth/abundance competition above; in this regime the practical bottleneck is not map noise but the availability of a large spin-orientation proxy catalog (Sec.\ \ref{sec:strategies}). The analytical and Monte-Carlo estimates (defined next) agree in Fig.\ \ref{fig:cum_snr}.

The cumulative SNR estimate of $\approx 68$ for the $10^{13}\,M_\odot$ mass-limited survey in Fig.\ \ref{fig:cum_snr} should be treated as an upper limit, as our fiducial Battaglia $n_e$ fits overestimate  $n_e$ when extrapolated to group-scale haloes \cite{qu26_ksz}; with this caveat, the mass range $10^{13}$-$10^{14}\, M_\odot$ contains a cumulative SNR of $\approx 43$ $(\sqrt{68^2-53^2})$ due to the abundance of group-scale haloes, even when the per-object SNR in Fig.\ \ref{fig:perobject_snr} is small.
Restricted to the validated range $\gtrsim10^{14}\,M_\odot$, the cumulative SNR is $\approx53$.

\subsection{Monte-Carlo estimation for Cumulative SNR}
In an observation, after obtaining our filtered vectors $\mathbf{V}_i$ or tensors $\mathbb{Q}_i$ ($i$ labels object), we can combine them into a single catalog-combined statistic using the weights defined in Eq.\ \eqref{eq:combine_weights}:
\begin{align}
	\hat{K}_\text{V}(\{\hat{\mathbf{d}}^\text{proxy}_i\}) &= \sum_i W_{i}^\text{V} \hat{\mathbf{d}}^\text{proxy}_i \cdot \mathbf{V}_i \label{eq:mc_v} \; , \\
	\hat{K}_\text{T}(\{e^\text{proxy}_{jk,i}\}) &= \sum_i W_{i}^\text{T} e^\text{proxy}_{jk,i} \mathbb{Q}^{jk}_i  \; , \label{eq:mc_t}
\end{align}
where $\hat{\mathbf{d}}^\text{proxy}_i$ are the projected rotation directions obtained from an external catalog.

To assess the SNR of this measurement, we can estimate the null distribution from the same data as follows: for each realization, we draw independent random isotropic directions $\hat{\mathbf{d}}_i^\text{rand}$ for each object, constructing a set of random unit vectors $\{\hat{\mathbf{d}}^\text{rand}_i\}$, and random axes  $\{e_{jk,i}^\text{rand}\}$ using Eq.\ \eqref{eq:dipole_tensor}. 
We therefore have numbers $\hat{K}_\text{V/T}^\text{rand}$ for each realization by substituting $\{\hat{\mathbf{d}}^\text{rand}_i\}$ and random axes  $\{e_{jk,i}^\text{rand}\}$ into Eqs.\ \eqref{eq:mc_v}--\eqref{eq:mc_t}. Across realizations, these numbers $\hat{K}_\text{V/T}^\text{rand}$ estimates the null distribution with variance
 \begin{align}
	 \sigma_\text{V/T,rand}^2 &= \operatorname{Var}(\hat{K}^\text{rand}_\text{V/T}) \; .
\end{align}
The Monte-Carlo cumulative SNR (for known dipole direction, $f_\text{misalign}=1$), to be compared with the analytical cumulative SNR is thus defined to be
\begin{align}
\text{cSNR}_\text{V,MC} &=  \frac{\hat{K}_\text{V}(\{\hat{\mathbf{d}}_i^\text{true}\})}{\sigma_\text{V,rand}} \\
\text{cSNR}_\text{T,MC} &=  \frac{\hat{K}_\text{T}(\{e_{jk,i}^\text{true}\})}{\sigma_\text{T,rand}} \; ,
\end{align}
where $\hat{\mathbf{d}}_i^\text{true}$ and $e_{jk,i}^\text{true}$  are the true rkSZ direction in the simulation.

\section{Discussion and Conclusion}
\label{sec:discussion}
\subsection{Modeling uncertainties on rkSZ template}
\label{sec:discuss_sim_diff}
We have derived a minimal, simulation-motivated rkSZ template: in the integrand we have a Battaglia generalized-NFW electron number density profile \cite{battaglia16_ne} multiplied by a radially constant equatorial speed $v_\mathrm{tan}=0.18\,V_{c,200c}$ (Eq.~\eqref{eq:vtan_to_vcirc}) and tapered by the shell--cluster misalignment profile $\bar\beta(x)$ (Eq.\ \eqref{eq:beta_model}). The model reproduces radial profiles $v_\text{tan}(x)$, $n_e v_\text{tan} \bar\beta (x)$, and integrated rkSZ statistic  $\mathcal{S}^\text{model}_{<x}$ observed in TNG-Cluster at different masses and redshifts (Figs.~\ref{fig:vtan_tng}, \ref{fig:nevtan_tng} and \ref{fig:Sx}).

The dominant modeling systematic of the rkSZ effect is the radial profile of the gas rotation, $v_\text{tan}(r)$, which shows variation across simulations in both its shape and amplitude. 
\begin{figure}
	\includegraphics[trim={0.0cm 0.0cm 0.0cm 0.75cm}, clip, width=\linewidth]{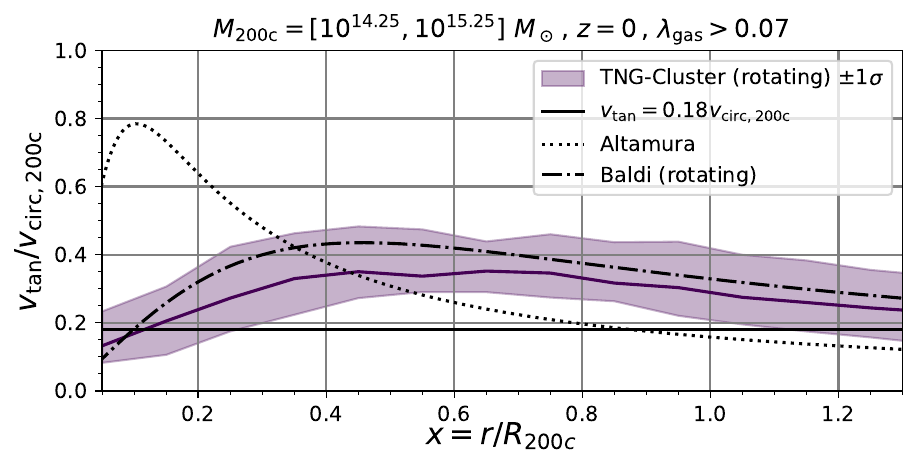}
	\caption{Tangential-velocity profiles $v_\text{tan}/V_\text{circ,200c}$ at $z=0$ for the TNG-Cluster \emph{rotating} subset with $\lambda_\text{gas}>0.07$ as in Baldi et al.\, (shaded band and line: 16th--84th percentile and median). The mass range of $10^{14.25}$ to  $10^{15.25}$  $M_\odot$ is chosen to be consistent with the mass ranges considered in Baldi et al.\ and Altamura et al. These are compared with our constant model (black), the Baldi et al.\ \cite{baldi17_rot} fast-rotator fit (dash-dotted; Eq.\ \eqref{eq:baldi_vtan}), and the Altamura et al.\ \cite{altamura23_ksz} all-sample fit (dotted; Eqs.\ \eqref{eq:altamura_omega}). 
\label{fig:fast_rotators}
}
\end{figure}

Here we compare the $v_\text{tan}$ profile of Altamura et al. \cite{altamura23_ksz} and Baldi's $v_\text{tan}$ profile for fast-rotating  $\lambda_\text{gas}>0.07$ subset of MUSIC clusters.
Altamura et al.\ \cite{altamura23_ksz} uses the following model for the angular velocity $\omega(r)$:
\begin{align}
	\omega(r) &= \frac{v_{t0}}{r_0 [1+(r/r_0)^\eta]} \; ,
	\label{eq:altamura_omega}
\end{align}
where $v_{t0}=1.53v_{c,500c}, \xi_0=10.02, r_0=0.16 r_{500c}, \eta=2.01$,  (all-sample fit, Table B2, Eq.\ (7) \cite{altamura23_ksz}).
The angular velocity $\omega(r)$ enters their Eq.\ (5), which is equivalent to our Eq.\ \eqref{eq:Theta0} with $r \omega \to \bar\beta v_\text{tan}$. Approximating $\eta\approx 2$, we recover the Baldi et al.\ parametrization for the fast-rotating subset of MUSIC clusters with $\lambda_\text{gas}>0.07$ (which differs by the same $\times3/2$ definitional factor from the Altamura et al.\ parametrization)
 \begin{align}
	 v_\text{tan}(r) = v_{t0} \frac{r/r_0}{1+(r/r_0)^2} \; .
	 \label{eq:altamura_vtan}
\end{align}
We take the Baldi et al.\ parameters from Table 4, \texttt{gas vp2b} for the AGN run.

\textbf{Shape.}
Our assumption $j(r)\propto r$ yields a flat $v_\mathrm{tan}(r)$, which captures the amplitude but not the radial, concave-down feature in TNG-Cluster (Fig.~\ref{fig:vtan_tng}). Baldi et al.~\cite{baldi17_rot} observe the same radial feature in its rotating and non-rotating sample (Fig.\ 6), and this is reflected by the fit $v_\mathrm{tan}(r)\propto (r/r_0)/[1+(r/r_0)^2]$ to the fast-rotating ($\lambda_\mathrm{gas}>0.07$) subset of the MUSIC clusters.
In Fig.\ \ref{fig:fast_rotators}, we compare the $v_\text{tan}/V_{c,200c}$ ratio extracted from TNG-Cluster $\lambda_\text{gas}>0.07$ sample with the Baldi fast-rotator $v_\text{tan}$ profile, our constant  $v_\text{tan}=0.18V_{c,200c}$ model, and Altamura's $z=0$ all-cluster $v_\text{tan}$ profile. 
The best-fit value for its angular velocity profile  $\omega(r)$ describes a shape and amplitude for $v_\text{tan}$ that is significantly different from our work and both Baldi's fast-rotating and non-rotating sample.
 
A fast-rotator sample selection changes the template in two ways--a larger amplitude $\mathcal{S}\propto\lambda_\mathrm{gas}$ (which affects the optimal object weighting for the combined SNR), and a different template shape $F(\ell)$ (which changes the matched-filter shape). If one knows \emph{a priori} (from the external proxy) whether the cluster is fast-rotating, one can replace the weights and filters to obtain a higher SNR than using the mean-population templates and weights. 

\textbf{Amplitude.}
While our simple model cannot capture the detailed shape of the $v_\text{tan}(r)$ profile, by construction its amplitude is consistent with numerically well-motivated values of the gas spin parameter  $\lambda_\text{gas}$, and agrees also with the amplitude of the non-rotating sample in Baldi et al.\ \cite{baldi17_rot} (Fig.\ 6, note our definition of $v_\text{tan}$ is $\times 3/2$ that of Baldi et al.). 
To check whether these numbers are sensible, we can substitute back into the spin parameter (Eqs.\ \eqref{eq:Ji_omegai}, \eqref{eq:avg_J}),
\begin{align}
	\lambda_\text{gas} &\approx \frac{2}{3\sqrt{2}} \frac{\int_0^1 n_e v_\text{tan} x^3 dx}{V_{c,200c} \int_0^1 n_e x^2 dx}  \; ,
	\label{eq:lambda_check}
\end{align}
Substituting the Battaglia profile into $n_e$ and taking $\bar\beta=1$ (or $\bar\beta$ in Eq.\ \eqref{eq:beta_model}), we have the spin parameter estimate  $\lambda_\text{gas}$ equal 0.044 (0.042) for our flat  $v_\text{tan}=0.18 V_{c,200c}$ model, 0.096 (0.091) for the Baldi et al.\  $v_\text{tan}$ model for fast-rotator sample ($\lambda_\text{gas}>0.07$), and 0.069 (0.063) for the Altamura $\omega$ model.
	The same estimates are obtained by evaluating using the Vikhlinin $n_e$ model \cite{vikhlinin06_ksz} (Table B1 ``All clusters'' row, Altamura et al.\ \cite{altamura23_ksz}). The estimate for Baldi's $v_\text{tan}$ profile is consistent with its fast-rotating subset. 
    We are unable to reconcile the quoted median spin parameter $\lambda_{500c,\text{gas}}=0.0505$ in Altamura et al.\ \cite{altamura23_ksz} with the spin parameter estimated with Eq.\ \eqref{eq:lambda_check}, modified to integrate to $R_{500c}$ instead of $R_{200c}$.

\textbf{Baryonic feedback.} Our signal template assumes the Battaglia
AGN-feedback profile for $n_e$~\cite{battaglia16_ne}, and our simulation
comparison uses IllustrisTNG (TNG-Cluster and TNG300-1). Stacked kSZ measurements around BOSS CMASS~\cite{schaan21,amodeo21_feedback} and DESI
luminous red galaxies~\cite{hadzhiyska25_ksz,berni25_ksz,qu26_ksz}, whose host
haloes have mean masses of
$\sim5\times10^{13}\,M_\odot$~\cite{amodeo21_feedback,berni25_ksz},
indicate that both the profile and simulation used in our work underpredict the effect of feedback at these masses. In other words, group-scale haloes have less gas lying within $R_{200c}$, which lowers $\mathcal S$, and the
dipole spreads to larger angular scales where the primary CMB dominates; both
effects lower the SNR relative to our forecast. As a simple estimate, if we take the effect of baryonic feedback to be an amplitude rescaling of the signal by a factor of 0.367 (as in Table III of Ref.\ \cite{qu26_ksz}) for all $M_{200c}\lesssim10^{14}\,M_\odot$, then the cumulative SNR for a mass-limited catalog down to $10^{13}\,M_\odot$ would be 55 instead of 68 in Fig.\ \ref{fig:cum_snr}.

\subsection{Foreground Dipoles}
Apart from the miscentering of clustered tSZ which is the dominant noise contribution to our estimator (and the clustered dust which we did not model in our analytical covariance but is present in the Monte-Carlo estimates), several other effects imprint a dipole-like pattern on cluster cutouts. Unless the dipole directions for these effects correlate with the external spin proxy, they only inflate the estimator variance without biasing a cross-correlation.

\textbf{The lensed CMB gradient}: the primary CMB is well-approximated as a large-scale gradient at angular scales of galaxy clusters; the cluster lenses the background primary CMB, generating a dipole parallel with the CMB gradient \cite{seljak00_cmbdipole}; the primary CMB gradient is uncorrelated with the cluster spin axis. 

\textbf{The moving-lens effect}:
	in the CMB frame, the transverse peculiar velocity $\mathbf{v}_\perp$ of the cluster corresponds to a time-dependent potential that induces an integrated Sachs-Wolfe effect, which produces a temperature dipole aligned with $\mathbf{v}_\perp$ \cite{hotinli26_movinglens}. 
	The large-scale peculiar velocity derives from the first derivative of the linear potential ($\mathbf{v}\propto\nabla\Phi$), whereas tidal-torque theory states that cluster spin is sourced until turnaround through the misalignment between the tidal tensor $\partial_j \partial_k \Phi$ and the protohalo inertia tensor $I_{jk}$ via $J_i \propto \epsilon_{ijk}I_{kl}\partial_j\partial_l\Phi$. The spin axis and the large-scale $\mathbf{v}_\perp$ should therefore be uncorrelated in linear theory.

Since the spin direction and other dipole signals are uncorrelated in the vector formalism, they average down in the spin-oriented vector cross-correlation. On the flip side, the data analysis formalism we developed above could be readily applied to produce the dipole fields for the lensed CMB, or the moving lens given the signal templates. It will be interesting to see whether our formalism complements existing works on other cosmologically-interesting dipoles.

\subsection{Detection Strategies}
\label{sec:strategies}
Our forecast shows that the tensor estimator, which manipulates the dipole into a quadrupolar signal for the purpose of cross-correlating with spin-2 fields, carries no statistical power for SO. We therefore focus exclusively on the vector estimator.

We can make a conceptual split of objects into the instrumental-dominated (ILC) regime, and the tSZ-dominated (cILC) regime, as seen in Fig.\ \ref{fig:perobject_snr}, corresponding to the noise component that dominates the diagonal of the covariance matrix (and the multifrequency cleaning method that works best in the regime). In practice, the multifrequency matched filter (MMF) accounts also for the non-negligible off-diagonal covariance from the clustered noise and proves to be optimal also in the transitional regime. We will stick to the two-regime split for its useful simplification.

\textbf{The tSZ-dominated (cILC) regime.} In this regime the optimal filter naturally converges to the tSZ-deprojected solution so as to remove the dominating tSZ power. The residual small-scale CIB and instrumental noise after deprojection determines the variance of the filter, $N_\text{V}^2$. The larger the residual small-scale power, the more of the filter weights will be shifted to large scales ($\ell\lesssim 3000$) where the CMB power dominates (see Fig.\ \ref{fig:Dk_comps}).
The detection strategy to be emphasized in this regime is multifrequency cleaning: not so much whether there exists a very low-noise single-frequency channel, but rather whether there are large enough numbers of frequency channels with comparable noise levels at small scales such that no large noise and CIB penalty will be incurred from deprojection.
It is also important to observe all highest-mass clusters, e.g.\ by increasing the survey sky fraction $f_\text{sky}$ or through targeted observation.
Future work should study the combination of multiple surveys, for example ACT or SPT, the CCAT-Prime 280, 350, 410 GHz \cite{ccat23} channels, as well as subarcminute resolution instruments such as NIKA2 \cite{adam18_nika2,catalano18_nika2} and TolTEC \cite{bryan18_toltec}. See Mroczkowski et al.\ \cite{mroczkowski19_szreview} for a comprehensive review.

\textbf{The instrumental-dominated (ILC) regime.} In this regime, the cluster-centered tSZ falls below the instrumental noise on the small angular scales, so the effective noise is approximately isotropic and the multifrequency matched filter converges to a single-frequency matched filter on ILC-cleaned maps (Fig.~\ref{fig:perobject_snr}). Deprojecting the tSZ here is counterproductive, since it pays a noise penalty to remove a subdominant component; a pipeline that applies tSZ deprojection uniformly across all masses therefore sacrifices roughly half of the cumulative SNR (Fig.~\ref{fig:cum_snr}).
For SO-goal specification, and an external proxy that traces the halo mass function, this regime contributes most of the SNR, as shown by the information density contours (Fig.~\ref{fig:info_den}). 
The detection strategy in this regime is different: it requires an external proxy catalog with a considerable fraction of clusters compared to a mass-limited survey at $M_{200c}\sim 10^{13.7}$--$10^{14.7}\,M_\odot$ and redshift of ${0.2\lesssim z \lesssim 1.0}$, as each object (of cosmological-average spin parameter) contributes $\sim 0.2 f_\text{misalign}$ SNR each, where $f_\text{misalign}$ is the misalignment between proxy and true spin direction (at least 25 clusters needed for SNR of 1). Therefore, unless a large spin-proxy catalog is accessible (current measurements rely on proxy catalogs which are small in size, e.g.\ $\lesssim 25$ rotating clusters from Manolopoulou \& Plionis \cite{manolopoulou17_galrot} in Goldstein et al.\ \cite{goldstein25_rksz} and Baxter et al.\ \cite{baxter19_rksz}), it will be difficult to extract all possible information within this regime. Another strategy that could be utilized is selectively picking fast rotators through the proxy, as done in Manolopoulou \& Plionis, so as to increase the combined SNR.

\textbf{Angular resolution.} As a clustered signal, the rkSZ template has considerable support at small scales $\ell\gtrsim 8000$ (especially for high-redshift clusters, see Fig.\ \ref{fig:1d_rksz}), beyond the current $\ell_\text{max}$ cutoff of this study.
A smaller beam would raise the maximum usable multipole and increase the single-object SNR for all mass and redshifts. It will be therefore interesting to study the rkSZ signal in combination with subarcminute instruments e.g.\ NIKA2 and TolTEC as discussed above.

When the size of the spin proxy catalog is limiting (as it is currently), the strategy may be to target the highest-mass clusters and go down in a mass-limited manner, and explore instrument combinations to optimize multifrequency cleaning and gain access to small-scale modes. The cumulative SNR we calculated in Fig.\ \ref{fig:cum_snr} would be the theoretical limit for SO-goal noise specification, if all objects have the cosmological-average spin parameter of $\lambda_\text{gas}$. Picking a fast-rotating subset scales the single-object SNR linearly with $\lambda_\text{gas}$ (via $\mathcal{S}\propto\lambda_\text{gas}$), at the cost of catalog size. Because $\lambda_\text{gas}$ follows a log-normal distribution, the heavy tail of the distribution may allow for the strategy of picking fast rotators: e.g.\ using the fiducial values of $\lambda_{0,\text{gas}}=0.046$ and $\sigma_\lambda=0.609$ from Ref.\ \cite{gottloeber07_spinsim}, then 12\% of the catalog should be contained in the fast-rotator sample with $\lambda_\text{gas}>2\lambda_{0,\text{gas}}\sim 0.09$.

\section{Conclusion}\label{sec:summary}
We have presented a formalism that turns microwave intensity maps into a projected gas angular-momentum field of galaxy clusters through the rotational kSZ effect, and forecasted its detectability with a realistic forward model including simulated SO-goal noise together with the lensed CMB, thermal SZ and the CIB.

The central object is a per-cluster two-dimensional vector $\mathbf{V}(\hat{\mathbf{n}})$ that estimates the rkSZ dipole direction $\hat{\mathbf{d}}=\hat{\mathbf{J}}\times\hat{\mathbf{n}}/|\hat{\mathbf{J}}\times\hat{\mathbf{n}}|$ and amplitude $\mathcal{S}$. The vector estimator is linear in the temperature map and is constructed through the minimum-variance unbiased weighting of harmonic modes at individual frequency channels, given a model of the signal profile and the noise covariance. It provides optimal scale separation compared to the real-space ``left-minus-right'' statistic used in previous oriented-stacking analyses \cite{baxter19_rksz,goldstein25_rksz,yang26_rksz}. When an external proxy supplies only the rotation axis and not its sign, we provide an analogous tensor estimator $\mathbb{Q}$ built from a quadrupolar manipulation of the same maps, though it carries no statistical power for current experiments.

We have introduced a simple and simulation-motivated model of the rkSZ which would be useful for future studies to estimate the size and shape of the mass-redshift dependent signal. It requires only a choice of the electron density profile, the gas spin parameter, and a simulation-calibrated spin misalignment profile $\bar\beta(x)$ between the total cluster and individual shells. Despite this simplicity, it matches the integrated rkSZ statistic $\mathcal{S}$ measured in TNG-Cluster+TNG300-1 to within the cluster-to-cluster scatter for $M_{200c}\gtrsim 10^{13.5} M_\odot$ (group and cluster scales).
While the flat shape of our modeled tangential velocity profile $v_\text{tan}$ cannot match the detailed concave-down shape observed in TNG-Cluster and in previous simulation work e.g.\ Baldi et al \cite{baldi17_rot} and Altamura et al.~\cite{altamura23_ksz}, our amplitude is in agreement with Baldi et al., and the shape of $v_\text{tan}$ remains a modeling uncertainty across simulations (Fig.\ \ref{fig:fast_rotators} and Sec.\ \ref{sec:discuss_sim_diff}).

Our forecast makes explicit that the optimal filtering for the rkSZ signal is not only scale-dependent, as noted by \cite{monllor24_szrot, zorrilla20_rksz}, but also mass dependent. We identified two distinct regimes that contribute to the detection---1. the tSZ-dominated (cILC) regime with the highest-mass ($\gtrsim 10^{14.7} M_\odot$) clusters and the highest single-object SNR, and 2.\ the instrumental-dominated (ILC) regime that carries the most cumulative SNR (for SO-goal specification and a mass-limited cross-correlation), as there is an abundant cluster population at $M_{200c}\sim 10^{13.7}$--$10^{14.7}\,M_\odot$ at $z\sim 0.2$--$1.0$; here the clustered tSZ is subdominant and the multifrequency matched filter reduces to a single-frequency matched filter on ILC-cleaned maps.
These two regimes require very different multifrequency cleaning approaches, generalized by our optimal multifrequency matched filter (MMF): deprojecting the tSZ in the instrumental-dominated regime is wasteful, and doing so uniformly across the catalog forfeits about half of the total signal-to-noise. Only the rare highest-mass clusters require tSZ deprojection, toward which the MMF filter converges naturally. 

For a mass-complete sample with $M_\text{min}=10^{14}M_\odot$ over $f_\mathrm{sky}=0.4$ with SO-goal noise and a perfect spin-direction proxy ($f_\mathrm{misalign}=1$), the cumulative vector SNR reaches $\approx 53$ with the multifrequency matched filter and $\approx 25$ with the simpler deprojected filter, while the axis-only tensor approach stays below unity.
Note these SNR numbers are idealized (as we have no access to a mass-limited spin proxy catalog), but act as a good theoretical limit of the statistical power one could achieve for a given experiment. Realistic forecasts would start from our single-object SNR estimate, and apply the mass-redshift distribution of the proxy catalog and the estimated misalignment $f_\mathrm{misalign}$.

We hope this work will contribute to the angular momentum field as a valuable cosmological observable in its own right, and to study alignments, clustering biases, and non-thermal pressure support. Cross-correlated with the reconstructed peculiar-velocity field it can also be used to build estimators sensitive to parity-asymmetry of initial conditions. For first detections of the rkSZ effect, our work contributes an optimal multifrequency cleaning and a reformulation of the existing oriented stacking technology; it formally introduces the angular momentum field of free electrons as a CMB observable, which can offer a new perspective on structure formation as SNR increases through an expanded spin proxy catalog and combined data from existing and next-generation microwave surveys and subarcminute instruments.

\section{Acknowledgments}

We thank Susmita Adhikari, Steve Allen, Richard Battye, Yan-Chuan Cai, Anthony Challinor, Jens Chluba, Alex van Engelen, Sam Goldstein, Colin Hill, Gil Holder, Yu-Hsiu Huang, Matt Jarvis, Andrew Pontzen, Eduardo Rozo, Andres Salcedo, Blake Sherwin, Kendrick Smith, Tianyi Yang, Yong Sheng Yap and Inigo Zubeldia for discussions.
This work received support from the U.S. Department of Energy under contract number DE-AC02-76SF00515 to SLAC National Accelerator Laboratory. Research at Perimeter Institute is supported in part by the Government of Canada through the Department of Innovation, Science and Economic Development and by the Province of Ontario through the Ministry of Colleges and Universities. JK acknowledges support from The Joseph Needham Foundation for Science \& Civilisation (Hong Kong) through the Joseph Needham PhD fellowship. WRC is supported by the Science and Technology Facilities Council (STFC) through an Ernest Rutherford Fellowship (UKRI2424: “Astrophysics and Cosmology from CMB Scattering”) and through the Royal Society grant ICA\textbackslash R2\textbackslash 252140.

This work was performed using resources provided by the Cambridge Service for Data Driven Discovery (CSD3) operated by the University of Cambridge Research Computing Service, provided by Dell EMC and Intel using Tier-2 funding from the Engineering and Physical Sciences Research Council (capital grant EP/T022159/1), and DiRAC funding from the Science and Technology Facilities Council.
\bibliography{references}

\appendix

\section{Analytical covariance modeling}
\label{app:cov}

The azimuthal-weighted covariance is defined as
\begin{align}
	\bar{R}_{\bar\ell \bar\ell'} &\equiv 
	 \int_{\bar\ell\text{-bin}} \frac{d^2\ell }{(2\pi)^2}\int_{\bar\ell'\text{-bin}} \frac{d^2 \ell'}{(2\pi)^2}\nonumber \\
				     &\phantom{ASDASD}\times\cos\phi \operatorname{Cov}(\operatorname{Im}[m(\Bell)],\operatorname{Im}[m(\Bell')])\cos\phi' \; .
\end{align}
We model the noise as two independent parts: a homogeneous isotropic part which leads to a diagonal covariance,
\begin{align}
	\bar{R}^\text{hom}_{\bar\ell\bar\ell'} &= \delta_{\bar\ell\bar\ell'} \bar{C}^\text{hom}_{\bar\ell} \; ,
\end{align}
and a clustered part from a miscenetered circularly-symmetric template $e^{i\Bell\cdot\mathbf{a}}y(\ell)$. The clustered part can be written as
\begin{align}
	\bar{R}^\text{clus}_{\bar\ell \bar\ell'} &\equiv 
	\int_{\bar\ell\text{-bin}} \frac{d^2\ell }{(2\pi)^2}\int_{\bar\ell'\text{-bin}} \frac{d^2 \ell'}{(2\pi)^2} \langle \sin(\Bell\cdot\mathbf{a})\sin(\Bell'\cdot\mathbf{a}) \rangle_{\mathbf{a}}  \nonumber \\
				     &\phantom{ASDASD}\times\cos\phi \langle y(\ell)y(\ell')\rangle \cos\phi' \nonumber \\
				     &\phantom{{}={}}-
				     \int_{\bar\ell\text{-bin}} \frac{d^2\ell }{(2\pi)^2}\int_{\bar\ell'\text{-bin}} \frac{d^2 \ell'}{(2\pi)^2} \langle \sin(\Bell\cdot\mathbf{a})\rangle_{\mathbf{a}} \langle\sin(\Bell'\cdot\mathbf{a}) \rangle_{\mathbf{a}}  \nonumber \\
				     &\phantom{ASDASD}\times\cos\phi \langle y(\ell)\rangle\langle y(\ell')\rangle \cos\phi' \; .
\end{align}
The miscentering vector $\mathbf{a}$ is assumed to follow a 2D Gaussian distribution centered at zero with a standard deviation $\sigma_a$:
\begin{align}
	\langle \sin(\Bell\cdot\mathbf{a})\sin(\Bell'\cdot\mathbf{a})  \rangle_{\mathbf{a}} &= e^{-\sigma_a^2(\ell^2+\ell'^2)/2}\sinh(\sigma_a^2\Bell\cdot\Bell') \\
	\langle \sin(\Bell\cdot\mathbf{a})  \rangle_{\mathbf{a}} &= 0 \; .
\end{align}
Therefore
\begin{align}
	\bar{R}^\text{clus}_{\bar\ell \bar\ell'} &=
	\int_{\bar\ell\text{-bin}} \frac{d^2\ell }{(2\pi)^2}\int_{\bar\ell'\text{-bin}} \frac{d^2 \ell'}{(2\pi)^2} \cos\phi \sinh(\sigma^2_a \Bell\cdot\Bell') \cos\phi'  \nonumber \\
						 &\phantom{ASDASD}\times \langle y(\ell)y(\ell')\rangle e^{-\sigma^2_a (\ell^2+\ell'^2)/2} \\
	&= \frac{1}{2}\int_{\bar\ell\text{-bin}} \frac{d^2\ell }{(2\pi)^2}\int_{\bar\ell'\text{-bin}} \frac{d^2 \ell'}{(2\pi)^2}  \nonumber \\
						 &\phantom{ASDASD}\times I_1(\sigma^2_a \ell\ell') \langle y(\ell)y(\ell')\rangle e^{-\sigma^2_a (\ell^2+\ell'^2)/2} \; ,
\end{align}
where $I_1$ is the modified Bessel function of order 1.
We approximate the integrand to be slowly varying and the circular profile has no variance
\begin{align}
	\bar{R}^\text{clus}_{\bar\ell \bar\ell'} &= 
	\frac{1}{2} I_1(\sigma^2_a \bar\ell\bar\ell') \bar{y}_{\bar\ell} \bar{y}_{\bar\ell'} e^{-\sigma^2_a (\bar\ell^2+\bar\ell'^2)/2} \; ,
\end{align}
where the binned circular profile is
\begin{align}
	\bar{y}_{\bar\ell} &= \int_{\bar\ell\text{-bin}} \frac{d^2\ell}{(2\pi)^2}\langle y(\ell)\rangle \;. 
\end{align}
For the tSZ profile, we compute $\bar{y}_{\bar\ell}$ from a fitting profile of the electron pressure  $P_e(x)$, where  $x=r/R_{200c}$, 
\begin{align}
	y_\text{3D}(x) &\equiv \frac{\sigma_T}{m_e c^2} P_e(x) \\
	y(\mathbf{x}_\perp) &=
	2\int_0^\infty dr_\parallel y_\text{3D}(r) \\
			    &= 2R_{200c} \int_{x_\perp}^\infty \frac{xdx }{\sqrt{x^2-x^2_\perp}} y_\text{3D}(x) \; .
\end{align}
The Fourier profile is
\begin{align}
	y(\Bell) &=
	4\pi \frac{R^3_{200c}}{D^2_A} \int_0^\infty x_\perp dx_\perp J_0(\tilde{k}x_\perp) \int_{x_\perp}^\infty \frac{xdx}{\sqrt{x^2-x^2_\perp}} y_\text{3D}(x) \\
	&= 4\pi \frac{R^3_{200c}}{D_A^2} \int_0^\infty x^2 dx y_\text{3D}(x) \frac{\sin(R_{200c}\ell x/D_A)}{R_{200c}\ell x/D_A} \; .
\end{align}
\subsection{Covariance: Tensor case}
We make the Gaussian approximation to the covariance for the tensor case,
\begin{align}
	\bar{R}_{\bar\ell\bar\ell'} 
	&=8 \int_{\bar\ell\text{-bin}} \frac{d^2\ell }{(2\pi)^2}\int_{\bar\ell'\text{-bin}} \frac{d^2 \ell'}{(2\pi)^2} \times \cos2\phi \cos2\phi' \nonumber \\
	&\phantom{ASDADSASAS}\times\operatorname{Cov}( \operatorname{Im}[m(\Bell)],\operatorname{Im}[m(\Bell')])^2 \; ,
\end{align}
In the case of isotropic homogeneous noise where  $\operatorname{Cov}(\operatorname{Im}[m(\Bell)],\operatorname{Im}[m(\Bell')])=(2\pi)^2C^\text{hom}_\ell[\delta^2(\Bell-\Bell')-\delta^2(\Bell+\Bell')]/2$, the covariance becomes diagonal:
\begin{align}
	\bar{R}_{\bar\ell\bar\ell'} 
	&= 2\times (2\pi)^2\delta^2(\mathbf{0}) \delta_{\bar\ell\bar\ell'}\int_{\bar\ell\text{-bin}} \frac{d^2 \ell}{(2\pi)^2} (C^\text{hom}_\ell)^2 \; .
\end{align}
With clustered noise, we can modify the covariance using the results above to get
\begin{align}
	\bar{R}^\text{clus}_{\bar\ell\bar\ell'} &= 
	8 \int_{\bar\ell\text{-bin}} \frac{d^2\ell }{(2\pi)^2}\int_{\bar\ell'\text{-bin}} \frac{d^2 \ell'}{(2\pi)^2} \langle y(\ell) y(\ell')  \rangle^2  \nonumber \\
						&\phantom{ASd}\times \cos2\phi \cos2\phi'  e^{-2\sigma_a^2(\ell^2+\ell'^2)/2}\sinh^2(\sigma_a^2\Bell\cdot\Bell') \\
	&\approx 2 \overline{yy}_{\bar\ell}\overline{yy}_{\bar\ell'} I_2(2\sigma_a^2\bar\ell\bar\ell')e^{-2\sigma_a^2(\bar\ell^2+\bar\ell'^2)/2} \; ,
\end{align}
where
\begin{align}
	\overline{yy}_{\bar\ell} &= \int_{\bar\ell\text{-bin}} \frac{d^2\ell }{(2\pi)^2} \langle y(\ell)  \rangle^2 \; .
\end{align}
There is an additional cross term,
\begin{align}
	\bar{R}^{\times}_{\bar\ell\bar\ell'}
	&= 16 \delta_{\bar\ell\bar\ell'}\int_{\bar\ell\text{-bin}}\frac{d^2\ell}{(2\pi)^2}C^\text{hom}_\ell \cos^2 2\phi \operatorname{Var}(\sin(\Bell\cdot\mathbf{a}) y(\ell)) \\
    &= 8 \delta_{\bar\ell\bar\ell'}\int_{\bar\ell\text{-bin}}\frac{d^2\ell}{(2\pi)^2}C^\text{hom}_\ell \langle y^2(\ell)\rangle e^{-\ell^2 \sigma^2_a } \sinh(\ell^2\sigma^2_a) \; .
\end{align}
and the total tensor covariance 
\begin{align}\bar{R}_{\bar\ell\bar\ell'}=\bar{R}_{\bar\ell\bar\ell'}^\text{hom}+\bar{R}^\text{clus}_{\bar\ell\bar\ell'}+\bar{R}^{\times}_{\bar\ell\bar\ell'} \; .
\end{align}

\subsection{Effect of Isotropic Miscentering}
If the cutout is miscentered w.r.t.\ the cluster center (which is assumed to coincide with the rotational dipole center) by miscentering displacement $\boldsymbol{\alpha}$ whose amplitude  $a$ follows a Rayleigh distribution with std $\sigma_a$, the mean Fourier profile modifies to
\begin{align}
	\bar{\Theta}^\text{miscen}_\text{rkSZ}(\boldsymbol{\theta})  &= 
	\langle \bar{\Theta}_\text{rkSZ}(\boldsymbol{\theta}-\boldsymbol{\alpha}) \rangle_{\boldsymbol{\alpha}}  \\
	\implies \bar{\Theta}^\text{miscen}_\text{rkSZ}(\Bell)  &= \langle e^{-i\Bell\cdot\boldsymbol{\alpha}}\rangle_{\boldsymbol{\alpha}} \bar{\Theta}_\text{rkSZ}(\Bell) \\
								&= -i\mathcal{S}\hat{\mathbf{d}}\cdot\hat{\Bell} e^{-\ell^2\sigma^2_a/2} F(\ell) \;, 
\end{align}
where $\langle \cdot  \rangle_{\boldsymbol{\alpha}}$ as the average over $\boldsymbol{\alpha}$. The rkSZ monopole profile would be modified to
\begin{align}
	F^\text{miscen}(\ell) &= e^{-\ell^2\sigma_a^2/2} F(\ell) \; .
\end{align}
\section{Constructing the vector matched filter}
\label{app:filter_vec}

Starting with the rkSZ component of the data vector $\mathbf{D}^\text{rkSZ}_{\bar\ell} =|\hat{\mathbf{J}}\times\hat{\mathbf{n}}| \mathcal{S} \bar{F}_{\bar\ell} \hat{\mathbf{d}} $ (Eq.\ \eqref{eq:for_appB_D}), the 2D vector $\mathbf{V} \equiv \sum_{\bar\ell} \Psi_{\bar\ell}\mathbf{D}_{\bar\ell}$ (Eq.\ \eqref{eq:for_appB_V}), and the covariance $\bar{R}_{\bar\ell \bar\ell'} \equiv \operatorname{Cov}(\hat{\Bell}_x\cdot\mathbf{D}_{\bar\ell},\hat{\Bell}_x\cdot\mathbf{D}_{\bar\ell'})$ (Eq.\ \eqref{eq:for_appB_cov}) where the direction $\hat{\Bell}_x$ is arbitrary. Since the filter $\Psi_{\bar\ell}$ is normalized such that
\begin{align}
    \langle \mathbf{V} \rangle = \langle |\hat{\mathbf{J}}\times\hat{\mathbf{n}}| \rangle \mathcal{S} \hat{\mathbf{d}} \; ,
\end{align}
the normalization constraint is 
\begin{align}
    \sum_{\bar\ell} \Psi_{\bar\ell} \bar{F}_{\bar\ell} = 1 \; .
\end{align}
Optimizing the variance $\operatorname{Var}(\mathbf{V}\cdot\hat{\mathbf{d}}) = \Psi_{\bar\ell}\bar{R}_{\bar\ell\bar\ell'} \Psi_{\bar\ell'}$ gives (Einstein summation implied)
\begin{align}
    &\partial_{\Psi_{\bar\ell}}\left[ \Psi_{\bar\ell'}\bar{R}_{\bar\ell'\bar\ell''} \Psi_{\bar\ell''} - N^2_\text{V} \bar{F}_{\bar\ell'} \Psi_{\bar\ell'}\right] = 0 \nonumber \\
    \implies& \Psi_{\bar\ell} = N^2_\text{V}\bar{F}_{\bar\ell'} \bar{R}^{-1}_{\bar\ell'\bar\ell} \; ,
\end{align}
where $N^2_\text{V}$ is a Lagrange multiplier. Substituting into the normalization constraint and variance,
\begin{align}
     N_\text{V}^{-2} &= \bar{F}_{\bar\ell'} \bar{R}^{-1}_{\bar\ell'\bar\ell} \bar{F}_{\bar\ell} \\
     \operatorname{Var}(\mathbf{V}\cdot\hat{\mathbf{d}}) &= N_\text{V}^4 \bar{F}_{\bar\ell'} \bar{R}^{-1}_{\bar\ell'\bar\ell} \bar{F}_{\bar\ell} = N^2_\text{V} \; .
\end{align}

In the case of homogeneous noise (e.g.\ instrumental noise) $n^\text{iso}(\Bell)$, we have $\bar{R}_{\bar\ell\bar\ell'} \equiv \delta_{\bar\ell\bar\ell'} \bar{C}^\text{hom}_{\bar\ell}$ (summation not implied; Eq.\ \eqref{eq:for_appB_Chom}), and $\bar{R}^{-1}_{\bar\ell\bar\ell'} \equiv \delta_{\bar\ell\bar\ell'} (\bar{C}^\text{hom}_{\bar\ell})^{-1}$. Substituting into the filter and variance yield Eqs.\ \eqref{eq:cilc_filter}--\eqref{eq:cilc_filter2}.

\subsection{Thermal SZ deprojection using multifrequency matched filters}
We employ a multifrequency data model
\begin{align}
	{m}^\mu(\Bell) = -i\mathcal{S} a^\mu \hat{\mathbf{d}}\cdot\hat{\Bell} F(\ell) + b^\mu y(\Bell) + n^\mu(\Bell) \; , 
\end{align}
where $a^\mu$ is the SED for rkSZ, and  $b^\mu$ is the SED for the component (tSZ) we wish to explicitly deproject.
We promote $\Psi_{\bar\ell}$ to  $\Psi^\mu_{\bar\ell} \equiv \Psi_{\bar{L}}$, with a frequency channel index  $\mu$, and a multi-index $\bar{L}\equiv(\mu,\bar\ell)$. Collecting the normalization condition  $\Psi_{\bar{L}} \bar{F}_{\bar{L}} =1$ (Eq.\ \eqref{eq:for_appB_multiD}), the multifrequency covariance $\bar{R}_{\bar{L}\bar{L}'}\equiv \operatorname{Cov}(\hat{\Bell}_x\cdot \mathbf{D}_{\bar{L}},\hat{\Bell}_x\cdot \mathbf{D}_{\bar{L}'}
)$, and the 2D vector $\mathbf{V}\equiv \Psi_{\bar{L}}\mathbf{D}_{\bar{L}}$ (Eq.\ \eqref{eq:for_appB_multiV}). The derivation follows the above with $\bar\ell\to\bar{L}$

\section{Extracting rkSZ properties from TNG-Cluster and TNG300-1}
\label{app:extract}
We use the gas cells of the 352 TNG-Cluster haloes at $z=0$, 0.5 and 1, and,
at $z=0$, of TNG300-1 haloes with $M_{200c}\geq10^{13.5}\,M_\odot$. For each
cell $j$ we use its mass $m_j$, its displacement $\mathbf{r}_j$ from the
cluster center (\texttt{GroupPos}, the potential minimum), its peculiar
velocity $\boldsymbol{v}_j$, and its electron abundance ($\mathtt{EA}_j$). The number of
free electrons in the cell is
\begin{align}
	N_{e,j} = \mathtt{EA}_{j}\,\frac{X_\mathrm{H}\,m_j}{m_p} \; ,
\end{align}
where $m_p$ is the proton mass and $X_\mathrm{H}=0.76$ the hydrogen
fraction.

\subsection{Shell quantities}
Following Sec.~\ref{sec:definitions}, the $i$-th shell (with central radius $r_i$, shell width $\Delta_r$ and volume $V_i$),
contains the $j$-th gas cells with $r_i-\Delta r/2<|\mathbf{r}_j|\leq r_i+\Delta r/2$
($j\in i$) and has volume $V_i$. Its free-electron mass, density and electron angular
momentum are
\begin{align}
	M_{e,i} &= \sum_{j\in i} m_e N_{e,j} \; , \qquad
	n_{e,i} = \frac{M_{e,i}}{m_e V_i} \; , \\
	\mathbf{J}_i &= \sum_{j\in i} m_e N_{e,j}\, \mathbf{r}_j\times\boldsymbol{v}_j \; .
\end{align}
The same sum over all cells within $R_{200c}$ gives $\mathbf{J}$, and hence
$\beta_i=\hat{\mathbf{J}}\cdot\hat{\mathbf{J}}_i$ (Eq.~\eqref{eq:beta_def}). Inverting the
solid-body relation of Eq.~\eqref{eq:Ji_omegai} gives the equatorial
tangential velocity of the shell,
\begin{align}
	\hat{v}_{\text{tan},i} \equiv \frac{3}{2}\, \frac{|\mathbf{J}_i|}{M_{e,i}\, r_i} \; ,
	\label{eq:tng_extract}
\end{align}
shown in Fig.~\ref{fig:vtan_tng}; the integrand of Fig.~\ref{fig:nevtan_tng}
is $\beta_i n_{e,i}\hat{v}_{\text{tan},i}$, computed as
\begin{align}
    n_{e,i} \hat{v}_{\text{tan},i} = \frac{3}{2 V_i r_i}\left|\sum_{j\in i} N_{e,j} \mathbf{r}_j\times\boldsymbol{v}_j\right| \; .
\end{align}
The factor $3/2$ converts the mean
specific angular momentum of a shell into its equatorial tangential velocity. Without this factor, one
obtains the quantity used by Baldi et al.~\cite{baldi17_rot},
$|\mathbf{J}_i|/(M_{e,i} r_i)$, which they compute with a slightly different
particle sum.

\subsection{Integrated statistics} 
For a line of sight
$\hat{\mathbf{n}}$ chosen to be perpendicular to the angular momentum vector $\hat{\mathbf{J}}$, the dipole direction
$\hat{\mathbf{d}}=\hat{\mathbf{J}}\times\hat{\mathbf{n}}$ is a unit vector.
With $\hat{\mathbf{r}}_{\perp,j}$ the unit vector of the projected position
$\mathbf{r}_{j,\perp} \equiv \mathbf{r}_j-(\mathbf{r}_j\cdot\hat{\mathbf{n}})\hat{\mathbf{n}}$, the
particle-sum versions of $\mathcal{S}^\text{model}_{<R}$ and
$\mathcal{D}^\text{model}_{<R}$ are
\begin{align}
	\mathcal{S}_{<r}(\hat{\mathbf{n}})\,D_A^2 &= -\frac{\sigma_T}{c} \sum_{|\mathbf{r}_j|<r} (\hat{\mathbf{d}}\cdot\hat{\mathbf{r}}_{\perp,j})\, N_{e,j}\, (\boldsymbol{v}_j\cdot\hat{\mathbf{n}}) \; , \\
	\mathcal{D}_{<r}(\hat{\mathbf{n}})\,D_A^2 &= -\frac{\sigma_T}{c} \sum_{|\mathbf{r}_j|<r} \mathrm{Sign}[\hat{\mathbf{d}}\cdot\hat{\mathbf{r}}_{\perp,j}]\, N_{e,j}\, (\boldsymbol{v}_j\cdot\hat{\mathbf{n}}) \; .
\end{align}
Both quantities $\mathcal{S}_{<r}D^2_A$ and $\mathcal{D}_{<r}D^2_A$ have units of area.

We evaluate both integrated statistics along five lines of sight
$\hat{\mathbf{n}}_1,\dots,\hat{\mathbf{n}}_5\perp\hat{\mathbf{J}}$, obtained by
rotating $\hat{\mathbf{n}}_1$ about $\hat{\mathbf{J}}$ in steps of $2\pi/5$, and
average,
\begin{align}
	\mathcal{S}_{<r}^{\mathrm{avg}} = \frac{1}{5}\sum_{k=1}^{5}\mathcal{S}_{<r}(\hat{\mathbf{n}}_k) \; ,
\end{align}
and likewise for $\mathcal{D}$, as we expect coherent rotation to contribute
equally to every line of sight perpendicular to $\hat{\mathbf{J}}$. Note the
average enters only the comparison with the simulations
(Figs.~\ref{fig:Sx} and~\ref{fig:montero_dorta}) and not the forecast, so to reduce the simulation scatter. 

\end{document}